\documentclass[10pt,journal,final,finalsubmission,twocolumn]{IEEEtran}
\usepackage{graphicx}  
\usepackage{subfigure} 
\usepackage{amssymb}
\usepackage{amsmath}   
\usepackage{url}
\usepackage{algorithmic}
\usepackage[boxed]{algorithm}

\newcommand{\ra}{\rightarrow}
\newcommand{\clambda}{\mbox{\boldmath{$\lambda$}}}

\newcommand{\cgamma}{\mbox{\boldmath{$\gamma$}}}
\newcommand{\cDelta}{\mbox{\boldmath{$\Delta$}}}
\newtheorem{theorem}{Theorem}
\newtheorem{remark}{Remark}

\newtheorem{scn}{Scenario}

\begin{document}
\title{Power Efficient Scheduling under Delay Constraints over Multi-user Wireless Channels}
\author{Nitin Salodkar,
        Abhay Karandikar,~\IEEEmembership{Member,~IEEE} and V. S. Borkar,~\IEEEmembership{Fellow,~IEEE}
% \thanks{Manuscript received January 20, 2002; revised August 13, 2002.}% <-this % stops a space
\thanks{Nitin Salodkar is with Department of Computer Science and Engineering, IIT Bombay, Powai, Mumbai India - 400076. Email: {nitins@cse.iitb.ac.in} }
\thanks{Abhay Karandikar is with Department of Electrical Engineering, IIT Bombay, Powai, Mumbai India - 400076. Email: karandi@ee.iitb.ac.in }
\thanks{V. S. Borkar is with the School of Technology and Computer Science, Tata Institute of Fundamental Research,
Homi Bhabha Road, Mumbai, India - 400005. Email: borkar@tifr.res.in }
}
\maketitle

\begin{abstract}
In this paper, we consider the problem of power efficient uplink scheduling in a Time Division Multiple Access (TDMA) system over a fading wireless channel. The objective is to minimize the power expenditure of each user subject to satisfying individual user delay. We make the practical assumption that the system statistics are unknown, i.e., the probability distributions of the user arrivals and channel states are unknown. The problem has the structure of a Constrained Markov Decision Problem (CMDP). Determining an optimal policy under for the CMDP faces the problems of state space explosion and unknown system statistics. To tackle the problem of state space explosion, we suggest  determining the transmission rate of a particular user in each slot based on its channel condition and buffer occupancy only. The rate allocation algorithm  for a particular user is a learning algorithm that learns about the buffer occupancy and channel states of that user during system execution and thus addresses the issue of unknown system statistics. Once the rate of each user is determined, the proposed algorithm schedules the user with the best rate. Our simulations within an IEEE 802.16 system demonstrate that the algorithm is indeed able to satisfy the user specified delay constraints. We compare the performance of our algorithm with the well known M-LWDF algorithm. Moreover, we demonstrate that the power expended by the users under our algorithm is quite low. 
\end{abstract}


\section{Introduction}
\label{sec:intro}
Broadband wireless networks like IEEE 802.16 \cite{wimax_std} and  3G cellular \cite{3gpp_site} are expected to provide Quality of Service (QoS) for emerging multimedia applications. One of the challenges in providing QoS is the time varying nature of the wireless channel due
to  multipath fading \cite{tse_book}. Moreover, for portable and hand-held devices, energy efficiency is also an important consideration.

In a multi-user wireless system, recent studies \cite{tse_hanly1,knopp} suggest that since the wireless
channel fades independently across different users, this diversity  can be exploited by \emph{opportunistically} scheduling
the user with the best channel gain. This leads to significant performance gain in terms of total system throughput. Such scheduling algorithms that
exploit the characteristics of the physical channel to satisfy some network level QoS performance metrics are refererred to as {\em cross layer}  scheduling algorithms \cite{cross_layer_review}. Power required to transmit \emph{reliably} at a certain rate under better channel conditions is much less than that required under poorer channel conditions  at the same rate \cite{tse_book}. This suggests that in order to save power, one should transmit at higher rates under better channel conditions, this leads to queuing delays. Moreover, since transmission power  is an increasing and strictly convex function of the transmission rate \cite{tse_book},  power efficiency can also be achieved by  transmitting the data at lower rates, albeit at the cost of increased queuing delay thus leading to a power-delay tradeoff. 

In this paper, we consider a single cell multi-user wireless uplink system with Time Division Multiple Access (TDMA). For such a system, we consider the problem of determining the user to be scheduled in each time slot so that the
average transmission power expended by \emph{each user} is minimized subject to a constraint on the average queuing delay
experienced by \emph{each user}. Moreover, we assume a peak power constraint, i.e., in each slot the transmission power of a user is less than or equal to a certain maximum value. This scenario may correspond to a base station scheduling users on an uplink in an IEEE 802.16 system  to satisfy delay constraint of each user.

There is a copious literature on cross layer scheduling algorithms. See \cite{berryreview} for a succinct review.  The scheduling problem is typically formulated as an optimization problem with an objective of efficiently allocating resources such as time, frequency bands, power, codes etc. to the users under physical layer (wireless channel)  and/or network layer QoS constraints. Various QoS constraints have been considered in the literature like system throughput, minimum rate, maximum delay, delay bound, queue stability and
fairness. A \emph{scheduling policy} is an allocation rule that allocates these resources based on parameters like channel conditions of the users, their queue lengths etc. In this paper, we concentrate on efficiently allocating power and rates to users based on their channel condition and queue length. The power allocation policy is considered feasible if it satisfies certain average or peak power constraints. On the other hand, the rate allocation policy is considered feasible if the physical layer can deliver the data  reliably to the users at a given rate. The set of all feasible rate tuples is called the feasible capacity region \cite{berryreview}. 

A scheduling policy is considered \emph{stable} if  the expected queue lengths are bounded under the policy. Many scheduling policies proposed in the literature have considered stability as a QoS criterion. In \cite{tse_hanly1}, the authors determine the throughput capacity region of a multi-access system, i.e., the set of all feasible rates with average power constraints. In \cite{yeh8}, the authors have shown that the throughput capacity region is same as the multi-access stability region (i.e., the set of all arrival vectors for which there exists some rate and power allocation policies that keep the system stable.). A scheduler is termed
\emph{thoughput-optimal} if it can maintain the stability of the system as long as the arrival rate is within the stability region. Thoughput optimal scheduling policies  have been explored in  \cite{tse_hanly1}, \cite{neely1}.   Longest Connected Queue (LCQ) \cite{tass_lcq} , Exponential (EXP) \cite{stolyar} , Longest Weighted Queue Highest Possible Rate (LWQHPR) \cite{yeh1}  and Modified Longest Weighted Delay First (M-LWDF) \cite{qos_shared} are other well known throughput optimal scheduling policies. In \cite{tse_hanly2}, the authors define the notion of \emph{delay limited capacity} of a multi-access system, i.e., the maximum rate achievable such that the delay is independent of the fading characteristics.

While thoughput-optimal scheduling policies maintain the stability of the queueing system, they do not necessarily guarantee small queue lengths and consequently lower delays. Delay-optimal scheduling deals with optimal rate and power allocation such that the average queue length and hence average delay are minimized for arrival rates within the stability region under average and peak power constraints.  Due to the nature of the constraints, there is no loss of optimality in choosing the rate and power allocation policies separately \cite{berryreview}. Hence to simplify the problem, one can choose any stationary power allocation policy that satisfies the peak and average power constraints. The delay optimal policy therefore deals with with optimal rate allocation for minimizing delays under a given power allocation policy. It has been shown that the Longest Queue Highest Possible Rate (LQHPR) policy  \cite{yeh_thesis} (besides being thoughput optimal) is also delay optimal for any symmetric power control under symmetric fading provided that the packet arrival process is Poisson and packet length is exponentially distributed.

Apart from throughput and delay optimal policies, opportunistic scheduling with various fairness constraints
have been explored in \cite{tse_channel,liu_shroff2}.

In this paper, our focus is on rate allocation with a constraint on peak power as well as average queueing delay which acts as the QoS metric. This problem for a single user wireless channel without the peak power constraints has been explored in the pioneering work of \cite{berry}. The problem with many generalizations on arrival  and channel gain processes have been considered in subsequent papers
\cite{wang_mand, goyal, mukul, krish, balaji, cruz_dp, gauresh}. In most of these papers, the scheduling
policy has been formulated as a control policy within the Markov Decision Process (MDP) framework. However, only
structural results of the optimal policy are available under various assumptions and that too for a single
user scenario only. There is very little work for extending the vast body of literature on delay constrained power efficient scheduling to multi-user scenario. Recently, in \cite{neely2}, the author has extended the asymptotic analysis of Berry-Gallger \cite{berry} for exploiting the power-delay tradeoff in multi-user system. The objective is to minimize the total power on the \emph{downlink} subject to user queue stability constraints. The author using the concept of Lyapunov Drift Steering has also given an algorithm that comes within a logarithmic factor of achieving the Berry-Gallager power-delay bound.  However, on the downlink, the base station typically transmits with a constant maximum power sufficient to reach the farthest user and hence power minimization in not a major concern. Moreover, minimizing the sum power can lead to unfairness, i.e., users with better average channel conditions might get a far higher share of the bandwidth than the users who have relatively poor average channel conditions. On the other hand, for the \emph{uplink}, the problem is to minimize the power of each user subject to individual delay constraint, which has not been addressed in the literature so far. 

In \cite{berry_thesis}, the author has extended the analysis for single user case to the multi-user case, albeit with only two users which can be applicable for the uplink also. Beyond two users, the problem becomes unwieldy to gain any useful insight, primarily due to large state space. For the two user case, the author has given an elegant near optimal policy where each user's rate allocation is determined by the joint channel states across users and the user's own queue state. Thus each user's queue evolution process behaves as if it were controlled by a single user policy. However, computation of user's transmission power still takes into account the joint channel and queue state processes.

Even for the single user case \cite{wang_mand, goyal, mukul, krish, balaji, cruz_dp, gauresh}, practical implementation of optimal policy is far from simple.  This is because  a knowledge of the probability distributions of the arrival and channel gain process is required for computing the optimal policy. This  knowledge is not available in practice. While, we have addressed this limitation by formulating an on-line algorithm within stochastic approximation framework in \cite{jsac_paper}, this algorithm still deals with the single user scenario. This algorithm does not assume any explicit knowledge of the probability distributions of the channel gain and arrival processes.
In this paper, we consider a multi-user wireless system. The {\em state} of the system is defined as the minimum information required by the scheduler for making scheduling decisions. For the multi-user scenario considered in this paper the state space is considerably large as compared to that for  the single user scenario. We illustrate this with a simple example. Let us assume that the channel condition of a user can be represented using $8$ states. This is a practical assumption and has been justified in \cite{chang1}. Let us assume that each user has a buffer in which at most $50$ packets can be stored.  For a single user system, the channel state and buffer occupancy of the user forms the state of the system in any time slot. The number of states is $8 \times 50 = 400$. Now consider a multi-user system with $4$ users. In this case, the state of the system consists of the channel state and buffer occupancy of each user. The state space consists of $50^4 \times 8^4 = 2.56 \times 10^{10}$ states. Furthermore, the number of states increases exponentially with the users. Hence  determining the optimal policy by estimating the dynamic programming value function would take prohibitively long time.  Hence in this paper, we propose a alternate approach.
  
In our approach, each user's queue evolution behaves as if it were controlled by a single user policy. Depending on
each user's channel state and queue size,  the algorithm  allocates a certain rate to each user in a slot using the single
user algorithm outlined in this paper. The algorithm then schedules the user with the highest rate in a slot. From the
structural properties of optimal policy for a single user scenario,  it is well known that the optimal policy is increasing in queue length and channel gain \cite{berry_thesis}. Thus more number of packets are transmitted when the queue length is greater or the channel gain is higher. Hence a user transmitting at a high rate has either very good channel condition, or large queue length. Scheduling such a user, therefore, either leads to its power savings or aids in satisfying the delay constraint.

The scheduling algorithms proposed in the literature like EXP scheduler \cite{stolyar}, LQHPR scheduler \cite{yeh_thesis}, M-LWDF \cite{qos_shared} scheduler require the queue length information for determining the scheduling decision. In the downlink scenario, this information is readily available to the scheduler residing at the base station. However, in the uplink scenario, this information needs to be communicated by the users to the scheduler. Communicating the queue length  information poses a significant  overhead. In our approach, each user determines the rate at which it would transmit if it were scheduled in a slot. All the users inform these rates to the base station. The base station then schedules the user with the highest rate. Thus by communicating the rates directly, we avoid the queue length communication overhead. 

The IEEE 802.16 system is an emerging system for broadband wireless access and is expected to provide QoS to the users.  Through our simulations in an  IEEE 802.16 system, we demonstrate that the algorithm is indeed able to satisfy the delay constraints of the users. Moreover, we demonstrate that the power expenditure of a user is commensurate with its delay requirement, the average arrival rate and average channel conditions. The higher the delay, lower the average arrival rate and better the average channel conditions, the lower is the power expenditure.

The contributions of this paper are summarized as follows:
\begin{enumerate}
\item We formulate the problem of minimizing the average power expended  by each user subject to a constraint on individual user delay as a constrained optimization problem. To the best of our knowledge, this problem has not been studied in a multi-user uplink scenario.
\item We propose an online algorithm that does not require the knowledge of the probability distributions of the channel states and the arrivals of the users.
\item The computational complexity of our approach increases only linearly with the number of users.
\item The communication overhead of our approach is  low and hence the algorithm is suitable for practical implementation. The algorithm satisfies the delay constraints of the users. We demonstrate the power efficiency our our algorithm through comparison with the M-LWDF algorithm within an IEEE 802.16 system simulation.
\end{enumerate}

The rest of the paper is organized as follows. In Section \ref{sec:sys_mod}, we present the system model. We formulate the problem as an optimization problem in Section \ref{sec:pf}, where we show that the problem has the structure of a Constrained Markov Decision Problem (CMDP). We discuss the issues like large state space and unknown system model in determining an optimal solution using the traditional CMDP solution techniques. In Section \ref{sec:te}, we consider and extension to the traditional single user scenario based on transmitter induced errors. In Section \ref{sec:algorithm}, we propose an online algorithm that is based on the extension to the single user scenario detailed in Section \ref{sec:te}. We also discuss the implementation issues. We present the simulation setup and discuss results in Section \ref{sec:ssr}. Finally, we  conclude in Section \ref{sec:concl}.

\section{System Model}
\label{sec:sys_mod}
\begin{figure}
\centering
\includegraphics[width=3in]{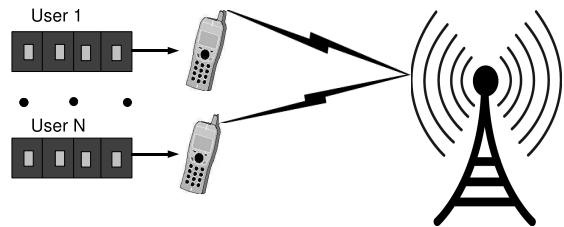}
\caption{System Model}
\label{fig:sys_mod}
\end{figure}
As illustrated in Figure \ref{fig:sys_mod}, 
we consider uplink transmissions in a TDMA system with $N$
users, i.e., time is divided into slots of equal duration and only one user is allowed
to transmit in a slot. We assume that the slot duration is normalized to unity. The base
station is a centralized entity that schedules the users in every slot. We assume a fading wireless
channel where the channel gain is assumed to remain constant for the duration of the slot and to change in an 
independent and identically distributed (i.i.d.) manner across slots. This model is called as the \emph{block fading} 
model \cite{berry}. We assume that the fading across users is also i.i.d. Under these assumptions,
if a user $i$ transmits a
signal $y^i_n$ in slot $n$, then the received signal $Y^i_n$ can be expressed as,
\begin{equation}
\label{eq:fading}
Y^i_n = H^i_n y^i_n + G_n,
\end{equation}
where $H^i_n$ denotes the complex channel gain due to fading and $G_n$ denotes the
complex additive white Gaussian noise  with zero mean and variance
$N_0$. Let $X^i_n=|H^i_n|^2$ be the channel state for user $i$ in slot
$n$. Practically $H^i_n$ is a continuous random variable and hence so is $X^i_n$. However,
in this paper we assume that $X^i_n$ takes only finite and discrete values from a set 
$\mathcal{X}$. This assumption has been justified in  \cite{berry}, \cite{wang_mand}. In this paper, we assume that the distribution of $H^i_n$ and hence that of $X^i_n$ is unknown. 

Each user possesses a finite buffer of $B$ bits. Bits arrive into the user buffer
and are queued until they are transmitted. The arrival process for each user is assumed to be i.i.d. across slots.
Let $A^i_n$ denote the number of bits arriving into the user $i$ buffer in slot $n$.  We assume that the random variable $A^i_n$ takes values from a finite and discrete set $\mathcal{A} \stackrel{\Delta}{=} \{0,\ldots,A\}$. Like $X^i_n$, we assume that the distribution of $A^i_n$ is  unknown.
Let $Q^i_n$ denote the queue length or buffer occupancy of user $i$ in slot $n$.  Let $U^i_n$ denote the number of bits transmitted by user $i$ in slot $n$. We assume that $U^i_n$ takes values from the set $\mathcal{U} \stackrel{\Delta}{=}
\{0,\ldots,B\}$. Let $I^i_n$ be an indicator variable that is set to $1$ if user $i$ is scheduled in slot $n$ and is set to $0$ otherwise. Let $\mathbf{I}_n$ be the vector $[I^1_n,\ldots,I^N_n]$. Note that since only one user can transmit in a slot, only one element of $\mathbf{I}_n$ is equal to $1$ and the rest are $0$. Let $\mathcal{I}$ be the set of all possible $N$ dimensional vectors with one element equal to $1$ and the rest being $0$. Let $K^i_n$ denote the number of bits that the user $i$ transmits in a slot if it is scheduled.  Then $U^i_n$ can be represented as $U^i_n = I^i_nK^i_n$. Moreover, since a user can at most transmit all the bits in its buffer in a slot, $K^i_n \leq Q^i_n$. Since we assume that the slot length is normalized to unity, $U^i_n$  is the rate at which user $i$ transmits in slot $n$. Let $\mathbf{U}_n$ be the vector $[U^1_n, \ldots, U^N_n]$, $\mathbf{U}_n \in \mathcal{U}^N$
% Let $\mathcal{U}^N$ be the set of all $N$ dimensional vectors such that at most one element of the vector is non-zero. The vector $\mathbf{U}_n$ takes values from the set $\mathcal{U}^N$.

The buffer evolution equation for user $i$ can be expressed as,
\begin{eqnarray}
\label{bufferevolution}
Q^i_{n+1}&=&\max \left\{Q^i_n-U^i_{n},0 \right \}+A^i_{n+1}.
\end{eqnarray}
The buffer size $B$ is large as compared to the arrival rate, thus we can neglect the buffer overflow in the buffer evolution equation. 

From \cite{berry}, the
power required for error-free or reliable communication at a rate $K^i_n = u$ bits/sec when $X^i_n=x$ is
given by,
\begin{equation}
\label{eq:power1}
   P(x,u) = \frac{W N_{0}}{x}\big(2^{\frac{u}{W} } -1\big).
\end{equation}
where $W$ is the bandwidth in Hz.
Note that for a given $x$, the transmission power $P(x,u)$ is an
increasing and  strictly convex function of  $u$. 
Let $\hat{P}$ denote the peak power constraint. Let $\hat{K}^i_n$ be the maximum rate at which user $i$ can transmit in a slot
$n$ when the channel condition is $X^i_n$ while satisfying the peak power constraint (i.e., $P(\hat{K}^i_n,X^i_n) \leq \hat{P}$).
Then the set of feasible rates for user $i$ in slot $n$, $\mathcal{F}^i_n \stackrel{\Delta}{=} \{0,\ldots,\min(\hat{K}^i_n,Q^i_n)\}$.

We assume that the users specify their QoS requirements in terms of the average
packet delay requirements. These delay requirements of the users are known
a priori to the scheduler. By Little's law \cite{gallager}, the average delay $\bar{D}$ is related
to the average queue length $\bar{Q}$ as,
\begin{equation}
\bar{Q} = \bar{a}\bar{D},
\end{equation}
where $\bar{a}$ is the average arrival rate. In the rest of the
paper, we treat average delay as synonymous with average queue
length and ignore the proportionality constant $\bar{a}$.

\section{Problem Formulation}
\label{sec:pf}
In this section, we formulate the problem as a constrained optimization problem within the Constrained Markov Decision Process (CMDP) framework. 
\subsection{Formulation as a Constrained Optimization Problem}
The user devices have limited battery power, hence it is essential to design transmission policies that conserve battery power. The power-delay tradeoff can be exploited to save power at the expense of extra delay. Moreover, multi-user diversity can be exploited to schedule a user with better channel state. Such a user requires lesser power while transmitting at a certain rate as compared to when it has a poorer channel state. However, this also incurs additional delay. The objective is to design a joint rate allocation and scheduling scheme that minimizes the power expenditure of each user subject to the satisfaction of the individual delay constraints.   
The average power consumed by a user $i$ over a long period of time can be expressed as,
\begin{eqnarray}
\label{eq:time_avg_power}
\bar{P}^i &=& \limsup_{M \rightarrow \infty}\frac{1}{M}\mathbf{E}\sum_{n=1}^{M}P(X^i_n,I^i_nK^i_n).
\end{eqnarray}
The average queue length of a user $i$ over a long period of time can be expressed as,
\begin{eqnarray}
\label{eq:time_av_q}
\bar{Q}^i &=& \limsup_{M \rightarrow \infty}\frac{1}{M}\mathbf{E}\sum_{n=1}^{M}Q^i_n.
\end{eqnarray}
Each user $i$ wants its average queue length to remain below a certain value, say, $\bar{\delta}^i$. Hence the problem becomes,
\begin{equation}
\label{eq:const_prob}
\mathrm{Minimize} \;\;\bar{P}^i \;\; \mathrm{subject\;to}\;\; \bar{Q}^{i} \leq \bar{\delta}^i,\;\; i = 1,\ldots,N.
\end{equation}
\begin{remark} {\em Dependence between problems:} Note that the $N$ problems formulated in (\ref{eq:const_prob}) are not independent. This is because in a TDMA system, only one user can be scheduled in a slot. Consequently, the scheduling decision in a slot impacts the buffer occupancy of all the users in the future slots.
\end{remark}
\subsection{Notion of Optimal Solution}
The problem in  (\ref{eq:const_prob})  is a multi-objective optimization problem with $N$ objectives and $N$ constraints. There can be multiple average power vectors that can be considered as optimal. Hence it is necessary to precisely define the properties of an optimal solution sought by us. We seek \emph{Pareto optimal} solutions \cite{multi_objective}. Let vector $[\bar{P}^1_{\psi},\ldots,\bar{P}^N_{\psi}]$ be the power expenditure vector under the rate allocation policy $\psi$. We say that the rate allocation policy $\psi$ is Pareto optimal if and only if there exists no rate allocation policy $\zeta$ with the corresponding power expenditure vector $[\bar{P}^1_{\zeta},\ldots,\bar{P}^N_{\zeta}]$ having the following properties,
\begin{eqnarray}
 \forall i \in \{1,\ldots,N\} P^i_{\zeta} \le P^i_{\psi} \;\;\land \;\;\exists i \in \{1,\ldots,N\}| P^i_{\zeta} < P^i_{\psi}.
\end{eqnarray}
The Pareto optimal solution is generally not unique and the set of Pareto optimal solutions is called the set of non-dominated solutions.
The weighted sum approach is a common approach for solving a multi-objective optimization problem \cite{multi_objective}. In this approach, one aggregates the $N$ objective functions into a single objective function. The resultant problem has a single objective function and $N$ constraints and can be expressed as,

\begin{eqnarray}
\label{eq:const_prob_single}
\mathrm{Minimize}\;\; \bar{P} &=& \gamma^1\bar{P}^1 +\ldots + \gamma^N\bar{P}^N, \nonumber \\
&&\mathrm{subject \; to},\nonumber \\
\bar{Q}^{i} &\leq& \bar{\delta}^i,\;\; i = 1,\ldots,N.
\end{eqnarray} 
where $\cgamma \stackrel{\Delta}{=} [\gamma^1,\ldots,\gamma^{N}]$ is the weight vector. It is generally assumed that $\gamma^i \in [0,1],\;\forall i,\;\sum_{i=1}^N\gamma^i =1$ implying that $\bar{P}$ is a convex combination of the individual powers. In general, the non-dominated set (i.e., the set of all Pareto optimal policies) may be a non-convex set. By varying the weight vector in  the weighted sum approach we can determine the Pareto optimal policies within a convex subset of the non-dominated set. However, choosing the weight vector in  order to obtain a particular solution is not straight forward. In the next section, we formulate the problem in (\ref{eq:const_prob_single}) within the CMDP framework.

\subsection{The CMDP Framework}
\label{subsec:cmdp}
Let $\mathbf{X}_n \stackrel{\Delta}{=}[X^1_n,\ldots,X^N_n]$ and $\mathbf{Q}_n \stackrel{\Delta}{=}[Q^1_n,\ldots,Q^N_n]$.
The state of the above system $\mathbf{S}_n$ at time $n$ can be described by
the tuple, $\mathbf{S}_n \stackrel{\Delta}{=} [\mathbf{Q}_n,\mathbf{X}_n]$, 
comprising of the queue length and the channel state of each of the $N$ users. 
Note that the system state space $\mathcal{S} = \mathcal{Q}^N \times \mathcal{X}^N$ is discrete and
finite. Let $\{\mathbf{S}_n\}$ denote the state process.
% A user $i$ is \emph{scheduled} in a slot $n$ if its rate $U^i_n$ in that slot is non-zero. For all the other users $j \ne i$ who are not scheduled in the slot $n$, the rate $U^j_n$ is set to zero. 
In each slot, the scheduler sets the rate vector  vector $\mathbf{U}_n \stackrel{\Delta}{=} [U^1_n,\ldots,U^N_n]$, where $U^i_n = I^i_nK^i_n$.  $\mathbf{U}_n$ takes values from the finite action space $\mathcal{U}^N$. $\{\mathbf{U}_n\}$ denotes the control process.  This problem has the structure of a CMDP with finite state and action spaces. Since we are considering average power expended and average delay suffered, it is an average cost CMDP. The scheduler objective is to determine an optimal  \emph{rate allocation policy}, i.e., a mapping from past history of states and actions to a rate allocation vector $\mathbf{U}_n$ in every slot $n$. For a CMDP with finite state and action spaces, it is well known that an optimal stationary randomized policy exists \cite{altman}, i.e., the rate allocation policy is a mapping from the current system state to a probability distribution on the set of rate allocation vectors. However, the traditional computational approaches based on Linear Programming \cite{altman} cannot be used to determine the optimal policy because of the following reasons:
\begin{enumerate}
\item \emph{Large state space:} The system state space is  large even for very few users. We have already illustrated this with an example in Section \ref{sec:intro}. Moreover, the state space grows exponentially with  number of users, hence the computational complexity of the traditional approaches also grows exponentially with number of users.
\item \emph{Unknown user/system statistics:} The probability distributions of $X^i_n$ and $A^i_n$ are unknown. The traditional approaches rely on the knowledge of these distributions for determining the optimal policy.
\end{enumerate}
We now extend the approach suggested in \cite{jsac_paper} to determine an optimal policy for the problem in (\ref{eq:const_prob_single}). 

% This algorithm is based on the algorithm developed in \cite{jsac_paper}. For the problem in (\ref{eq:const_prob_single}), the objective is to minimize the weighted sum power subject to individual user delay constraints. The state of the system is characterized by the tuple consisting of the channel state and buffer occupancy of all the users. Let $\mathbf{S}_n = [Q_n,X_n]$ denote the state of the system in slot $n$. Let $\mathbf{U}_n=[u^1_n,\ldots,u^N_n]$ be the rate vector such that the user who is scheduled has a non zero rate, and all the other users have zero rate.
Let user $i$ be scheduled in  slot $n$. Then the state of the system immediately after user $i$ transmits the data can be represented as $\mathbf{\tilde{S}}_n = [\mathbf{\tilde{Q}}_n, \mathbf{\tilde{X}}_n] =  [(q^1_n,\ldots,(\max(0,q^i_n-u^i_n),\ldots,q^N_n,x^1_n,\ldots,x^N_n]$. Let $\mathbf{A}_{n} = [a^1_n,\ldots,a^N_n]$ denote the arrival vector in slot $n$.  The state of the system at the beginning of slot $n+1$ can be represented as,
$\mathbf{S}_{n+1} = [\mathbf{Q}_{n+1},\mathbf{X}_{n+1}] = [(q^1_n+a^1_{n+1},\ldots,(\max(0,q^i_n-u^i_n+a^i_{n+1}),\ldots,q^N_n+a^N_{n+1},x^1_{n+1},\ldots,x^N_{n+1})]$.  
The queue transition equation in the vector form can be written as,
\begin{equation}
\label{eq:state_tran_vec}
\mathbf{Q}_{n+1} = \max(0,\mathbf{Q}_n + \mathbf{A}_{n+1} - \mathbf{U}_n).
\end{equation} 
Let $\cDelta = [\bar{\delta}^1,\ldots,\bar{\delta}^N]$ denote the delay constraint vector.
The problem in (\ref{eq:const_prob_single}) can be converted into an unconstrained problem using the Lagrangian approach. The unconstrained problem can be expressed as,
\begin{equation}
\label{eq:opt_uncons}
\mathrm{Minimize}\;\; \bar{P} + \sum_{i=1}^{N} \lambda^{i} (\bar{Q}^i - \bar{\delta}^i). 
\end{equation} 
The immediate cost $b(\cdot,\cdot,\cdot)$ incurred in scheduling a user $i$ in state $\mathbf{S}_n$ when the LM vector is $\clambda$ and rate vector $\mathbf{U}_n$ (user $i$ is scheduled in slot $n$) can be expressed as,
\begin{equation}
b(\clambda,\mathbf{S}_n,\mathbf{U}_n) = P(x^i_n,u^i_n) +\sum_{i=1}^{N} \lambda^{i} (Q^i_n - \bar{\delta}^i)
\end{equation} 
Let $\mathbf{\tilde{s}}^{0} = [\mathbf{q}^0,\mathbf{x}^0]$ denote a fixed state. Let $\tilde{V}(\cdot)$ denote the dynamic programming value function based on the state $\mathbf{\tilde{S}}$ reached immediately after taking the scheduling decision but before the arrivals. Let $\mathcal{F}_n = [\mathcal{F}^1_n,\ldots,\mathcal{F}^N_n]$ be the set of feasible\footnote{A rate is feasible based on the peak power constraints and the buffer occupancy} rate vectors  in slot  $n$.
Let $\{f_n\}$ and $\{e_n\}$ be two sequences that have the following properties,
\begin{eqnarray}
\label{eq:f_properties1}
f_n \rightarrow 0, \quad e_n \rightarrow 0, \quad \sum_n (f_n)^2 < \infty, \quad \sum_n (e_n)^2 < \infty, \\
\label{eq:f_properties2}
\sum_n f_n = \infty, \quad \sum_n e_n = \infty,   \\
\label{eq:f_properties3}
\quad \sum_n ({f_n}^2 + {e_n}^2) < \infty, \quad \lim_{n \rightarrow \infty} \frac{e_n}{f_n} \to 0.
\end{eqnarray}
The significance of these properties is explained later. We now present an optimal online primal-dual algorithm for solving the constrained problem in (\ref{eq:const_prob_single}):
\begin{eqnarray}
\label{eq:rate_deter7} 
\mathbf{U}_{n+1} &=& \arg\min_{\mathbf{V} \in \mathcal{F}_n}\Big\{(1-f_n)\tilde{V}(\mathbf{\tilde{S}}_n) +  f_n \times \nonumber \\
&&\big\{b(\clambda,(\mathbf{\tilde{Q}}_n+ \mathbf{A}_{n+1}, \mathbf{X}_{n+1}),\mathbf{V})   \nonumber \\
&&+\tilde{V}_n((\mathbf{\tilde{Q}}_n+ \mathbf{A}_{n+1} - \mathbf{V},\mathbf{X}_{n+1}))  \nonumber \\
&&-\tilde{V}_n(\mathbf{\tilde{s}}^{0})\big\}\Big\}, \\
\label{odprvi7} \tilde{V}_{n+1}(\mathbf{\tilde{S}}_n) &=& 
(1-f_n)\tilde{V}_{n}(\mathbf{\tilde{S}}_n) +  f_n \times \nonumber \\
&&\Big\{b(\clambda,(\mathbf{\tilde{Q}}_n+ \mathbf{A}_{n+1}, \mathbf{X}_{n+1}), \mathbf{U}_{n+1}) \nonumber \\
&&  +\tilde{V}_n(\mathbf{\tilde{Q}}_n+ \mathbf{A}_{n+1} - \mathbf{U}_{n+1}, \mathbf{X}_{n+1})  \nonumber \\
&&- \tilde{V}_n(\mathbf{\tilde{s}}^{0})\Big\}.
\end{eqnarray}
The algorithm in (\ref{eq:rate_deter7}) and (\ref{odprvi7}) is an online version of the well known Relative Value Iteration Algorithm (RVIA) \cite{ber}. It iteratively determines the optimal value function and hence the optimal policy one state at a time for a fixed value of the LM vector $\clambda$. To determine the optimal LM vector we augment the above algorithm with a dual LM iteration:
\begin{equation}
\label{eq:opt_lm}
\clambda_{n+1} = \Lambda[\clambda_{n} +e_n\left(\mathbf{Q}_n-\cDelta\right)],
\end{equation} 
where $\Lambda$ is a projection operator for ensuring that the LMs are non-negative and finite.
The properties of the update sequences in (\ref{eq:f_properties2}) ensure that the  sequences $\{f_n\}$ and $\{e_n\}$ converge to $0$ sufficiently fast to eliminate the noise effects when the iterates are close to their optimal values $\tilde{V}(\cdot,\cdot)$ and $\clambda^{*}$,  while those in (\ref{eq:f_properties1}) ensure that they do not approach $0$ too rapidly to avoid convergence of the algorithm to non-optimal values. Furthermore, (\ref{eq:f_properties3})  ensures that the update rates of primal iterations, i.e., the value function iterations and the dual iterations, i.e., the LM iterations are different. Since $e_n$ approaches $0$ much faster than $f_n$, the update rate of the value function iterations is much higher than the update rate of the LM iterations. This ensures that even though both the primals and duals are updated simultaneously, both converge to their optimal values \cite{bor_two_time}.
(\ref{eq:rate_deter7}),  (\ref{odprvi7}) and (\ref{eq:opt_lm}) constitute the optimal algorithm. The proof of optimality is exactly similar to that  in \cite{jsac_paper}. 

However, compared to the single user case, the state space here is too large for the algorithm to converge in reasonable number of iterations. We therefore motivate an alternate approach. We incorporate the possibility of transmitter induced errors in the single user scenario. We then motivate the multi-user solution by making use of this extension to the single user scenario.

\section{Single User Scenario in Presence of Transmitter Errors}
\label{sec:te}
We describe the scenario in brief. Consider a point to point transmission system over a fading wireless channel. Time is divided into slots of unit duration. We consider the block fading model as described in Section \ref{sec:sys_mod}. The scheduler is unaware of the probability distribution of the arrivals and the channel state in each slot. The objective is to minimize the average transmission power subject to average packet delay constraints. In \cite{jsac_paper}, we have determined an online algorithm that determines the optimal transmission rate in each slot so as to minimize the average power expenditure subject to average packet delay constraints. We consider the following extension to the above problem: suppose that after the online algorithm has determined the rate $U_n \in \mathcal{F}_n$, with a certain unknown random probability $\theta_n \in [0,1]$, the transmitter is unable to proceed with the transmission. We assume that the probability distribution of $\theta_n$ is not known. Under this assumption, the queue evolution equation can be expressed as,
\begin{equation}
\label{eq:q_ev_mod}
	Q_{n+1} = Q_{n} + A_{n+1} - I_n U_n, 
\end{equation} 
where $I_n$ is an indicator variable that is set to $1$ if the transmitter is successful in transmitting the packets and is set to $0$ otherwise. 
We now formulate the  rate allocation problem for this scenario. The long term power expenditure can be expressed as,
\begin{eqnarray}
\label{eq:time_avg_power_decom}
\bar{P}_e &=& \limsup_{M \rightarrow \infty}\frac{1}{M}\mathbf{E}\sum_{n=1}^{M}P(X_n, I_n U_n)
\end{eqnarray}
The average queue length over a long period of time can be expressed as,
\begin{eqnarray}
\label{eq:time_av_q_decom}
\bar{Q}_e &=& \limsup_{M \rightarrow \infty}\frac{1}{M}\mathbf{E}\sum_{n=1}^{M}Q_n,
\end{eqnarray}
Hence the rate allocation problem can be stated as,
\begin{equation}
\label{eq:const_prob_decom}
\mathrm{Minimize} \;\;\bar{P}_e \;\; \mathrm{subject\;to}\;\; \bar{Q}_e \leq \bar{\delta}.
\end{equation}
Note that the problem in (\ref{eq:const_prob_decom}) has the structure of a CMDP  with a state space for the single user case and average cost criterion. The objective is to determine an optimal policy $\mu^*$ such that the power expended under this policy is minimum possible while satisfying the delay constraint.

\subsection{The Primal Dual Approach}
The constrained problem in (\ref{eq:const_prob_decom}) can be converted into an unconstrained problem using the Lagrangian approach \cite{altman}. Let $\lambda \geq 0$ be a real number called as the Lagrange Multiplier (LM). Let $\mathcal{B}$ be the set $\{0,1\}$. $c:\mathcal{R}^+\times\mathcal{Q}\times\mathcal{X}\times\mathcal{B}\times\mathcal{U} \rightarrow \mathcal{R}$ be defined as the following,
\begin{eqnarray}
 c(\lambda,Q_n,X_n,I_n,U_n) \stackrel{\Delta}{=} P(X_n,I_nU_n) + \lambda(Q_n - \bar{\delta}),
\end{eqnarray} 
where $U_n$ is determined using the rate allocation policy $\mu:\mathcal{Q}\times\mathcal{X} \ra \mathcal{U}$.
The unconstrained problem is to minimize,
\begin{equation}
 \label{eq:uncons}
	L(\mu,\lambda) = \limsup_{M \rightarrow \infty}\frac{1}{M}\sum_{n=1}^{M}c(\lambda,Q_n,X_n,I_n,\mu(Q_n,X_n)).
\end{equation} 
$L(\cdot,\cdot)$ is called the Lagrangian. Our objective is to determine the optimal rate allocation policy $\mu^{*}$ and optimal LM $\lambda^*$ such that the following saddle point optimality condition is satisfied,
\begin{equation}
 \label{eq:saddle}
	L(\mu^{*},\lambda) \leq L(\mu^{*},\lambda^{*}) \leq L(\mu,\lambda^{*}).
\end{equation}  

For a fixed LM $\lambda$, the problem is an unconstrained Markov Decision Problem (MDP) with finite state and
action spaces with the average cost criterion. The following dynamic programming equation \cite{puterman} gives a necessary condition for optimality of a solution, 
\begin{eqnarray}
\label{eq:bellman}
V(q,x) = \min_{r \in \mathcal{F}}\Big[ c(\lambda,q,x,I,r)- \beta + \nonumber \\ \sum_{{a}',{x}'}p((q,x),r,(q+{a}'-r,{x}')) \times \nonumber \\
V(q+{a}'-r,{x}') \Big], {a}' \in \mathcal{A},\; {x}' \in \mathcal{X},
\end{eqnarray}
where $V(\cdot,\cdot)$ is the value function, $\beta \in {\mathcal{R}}$ is the unique optimal power expenditure.  Let $(q^{0},x^{0}) \in \mathcal{Q} \times \mathcal{X}$ be a fixed state. If we impose $V(q^{0},x^{0}) =0$, then $V(\cdot,\cdot)$ is unique \cite{puterman}. $p(s,r,s')$ is the probability of reaching a state ${s}'$ upon taking an action $r$ in state $s$. The traditional approaches for computing the optimal policy for an unconstrained average cost MDP such as the Relative Value Iteration Algorithm \cite{puterman} require the knowledge of $p(\cdot,\cdot,\cdot)$ which in this case is dependent on the probability distributions of the arrivals and channel states which is not known.
Note that determining the optimal value function as defined in (\ref{eq:bellman}) is not sufficient because the unconstrained solution for a particular $\lambda$ does not ensure that the constraints would be satisfied. To ensure constraint satisfaction, the optimal LM needs to be determined.

\subsection{The Online  Rate Allocation Algorithm}
We now present the rate allocation algorithm.  Let the user state at the beginning of slot $n$ be $(Q_n,X_n) = (q, x)$. Suppose that  $u$ bits are transmitted in slot $n$. The following primal-dual algorithm can be used to compute the rate $U_{n+1}=r_{n+1}$ at which the transmitter should transmit in slot $n+1$,
\begin{eqnarray}
\label{eq:rate_deter} 
r_{n+1} &=& \arg\min_{v \in \mathcal{F}_{n+1}}\Big\{(1-f_n)\tilde{V}_{n}(\tilde{q},\tilde{x}) +  f_n \times \nonumber \\
&&\big\{c(\lambda_n,\tilde{q}+ A_{n+1}, X_{n+1},1,v) \nonumber \\
&&+\tilde{V}_n(\tilde{q}+ A_{n+1} - v, X_{n+1}) \nonumber \\
&&-\tilde{V}_n(\tilde{q}^{0},\tilde{x}^{0})\big\}\Big\}, \\
\label{odprvi6} \tilde{V}_{n+1}(\tilde{q},\tilde{x}) &=&
(1-f_n)\tilde{V}_{n}(\tilde{q},\tilde{x}) +  f_n \times \nonumber \\
&&\Big\{c(\lambda_n,\tilde{q}+ A_{n+1}, X_{n+1},I_{n+1},r_{n+1}) \nonumber \\
  &&  +\tilde{V}_n(\tilde{q}+ A_{n+1} - I_{n+1}r_{n+1}, X_{n+1}) - \nonumber \\
&&\tilde{V}_n(\tilde{q}^{0},\tilde{x}^{0})\Big\},  \\
\label{eq:lm_iter6} 
\lambda_{n+1} &=& \Lambda[\lambda_{n} +e_n\left(Q_n-\bar{\delta}\right)].
\end{eqnarray}
These equations are explained below:
\begin{enumerate}
\item  (\ref{eq:rate_deter}), (\ref{odprvi6}) and (\ref{eq:lm_iter6}) constitute  the rate allocation algorithm.  It consists of two phases: \emph{rate determination phase} and \emph{update phase}. (\ref{eq:rate_deter}) constitutes the rate determination phase of the algorithm, i.e., it is used to determine the rate at which a user transmits in a slot \emph{if the transmission is successful}. (\ref{odprvi6}) is the primal iteration to determine the optimal value function and thereby the optimal policy, while (\ref{eq:lm_iter6}) is the coupled dual iteration for determining the optimal LM.  They constitute the update phase of the  algorithm.
\item If in a state $(Q_n,X_n) = (q, x)$, the transmitter decides to transmit $u\le q$ bits, then $\tilde{q} \stackrel{\Delta}{=} q-u$, and $\tilde{x} \stackrel{\Delta}{=} x$.
\item (\ref{odprvi6}) determines the optimal value function based on this new virtual state $(\tilde{q},\tilde{x})$. Note that the value function for this new state is related to the usual value function as  $\tilde{V}_n(\tilde{q},\tilde{x})  = \mathbf{E}^{A,X}[V_n(Q,X)]$.
% \item The expectation has been dropped and averaging performed using stochastic approximation \cite{spall} to arrive at iteration (\ref{odprvi6}). Note that in (\ref{odprvi6}), the value function is updated for the state $(\tilde{q}^i,\tilde{x}^{i})$ only; for the rest of the states, the value function remains the same. 
\item 
% Note that the immediate cost function $c(\cdot,\cdot,\cdot,\cdot,\cdot)$ in (\ref{eq:rate_deter}) has $1$ as its fourth argument, i.e., $I^i_{n+1}=1$. 
The rate determination phase (\ref{eq:rate_deter}) determines the rate  assuming that the transmitter would be successful in transmitting in slot $n+1$ ($I_{n+1}$ is assumed to be equal to $1$). 
However in (\ref{odprvi6}),  updating the value function requires the knowledge of whether the transmission is successful or not. This because the immediate cost function $c(\cdot,\cdot,\cdot,\cdot,\cdot)$ depends on $I_{n+1}$, i.e., whether the transmission is successful or not. Thus the update phase updates the value function and LM in each slot based on the success of the transmission.
\item $(\tilde{q}^{0},\tilde{x}^0)$ is any pre-designated state. On the RHS in (\ref{odprvi6}), the value function corresponding to this state is subtracted in order to keep the iterates bounded.
\item The LM iteration in (\ref{eq:lm_iter6}) ensures that the specified delay constraint is satisfied.
\item The sequences $f_n$ and $e_n$ have properties specified in (\ref{eq:f_properties1}), (\ref{eq:f_properties2}) and (\ref{eq:f_properties3}). The reasons for imposing these properties have been explained in Section \ref{subsec:cmdp}.
\end{enumerate}

\subsection{Proof  of Convergence}
\begin{theorem}
For the rate determination algorithm (\ref{eq:rate_deter}), (\ref{odprvi6}) and (\ref{eq:lm_iter6}), the  iterates $(V_n,\lambda_n ) \ra (V,\lambda^*)$. 
\end{theorem}
\begin{proof}
The proof of convergence is exactly similar to that in \cite{jsac_paper}. The probability of transmission failure in each slot serves as an extra \emph{noise} term. The algorithm being a stochastic approximation based online algorithm, averages out this extra noise term and determines the optimal policy and the optimal LM.
\end{proof}

\section{An Online Primal Dual Algorithm For the Multi-user Problem}
\label{sec:algorithm}
In this section, we propose a suboptimal approach to solve the problem in (\ref{eq:const_prob}).  
% The optimal algorithm determines the rate $K^i_n \in  \mathcal{F}^i_n$ at which user $i$ transmits in a slot $n$, if it is scheduled, based on the entire state of the system $\mathbf{S}_n$.  
To avoid the state space explosion, in the proposed algorithm, we determine the rate $R^i_n \in \mathcal{F}^i_n$ for a user $i$  in a slot $n$, \emph{if it is scheduled}, based on its \emph{state} $S^i_n \stackrel{\Delta}{=} [Q^i_n,X^i_n]$ \emph{alone} instead of the entire system state $\mathbf{S}_n = [\mathbf{Q}_n,\mathbf{X}_n]$. Note that $S^i_n \in \mathcal{Q}\times \mathcal{X}$. The rate $R^i_n$ is determined using a rate allocation policy $\rho^i$, i.e., a mapping from the history of states and rate allocations for user $i$ to its transmission rate. Once the rate $R^i_n$ for each user $i$ is determined, the next task is to determine a  user to be scheduled in that slot.  The user selection policy $\kappa$ is a mapping, $\kappa:\mathcal{F}^1_n \times\ldots \times \mathcal{F}^N_n\rightarrow \mathcal{I}$.

\subsection{Rate Allocation Algorithm for a User}
The rate allocation algorithm for each user behaves as if it were controlled by a single user policy in the presence of transmitter errors as explained in Section \ref{sec:te}. Each user $i$ determines the rate $R^i_{n+1}$ at which it would transmit in slot $n+1$ if it were to be scheduled in slot $n+1$ and informs this rate to the base station. The base station uses the user selection algorithm to schedule a user. The users who are not scheduled in a slot update their value function assuming transmitter errors,  while the user who is scheduled updates its value function assuming successful transmission. 
\begin{remark}
In the case of the single user scenario with transmitter errors, the probability with which a transmission is unsuccessful is independent of the scheduler action, i.e., the transmission rate determined by the online algorithm. In the multi-user scenario, this independence does not hold. This makes the problem a multi-agent learning problem \cite{young_book}, \cite{young_poss} where each agent (user) attempts to learn the optimal strategy and the actions taken by an agent (a user)  influences the actions taken by the other agents (users). 
\end{remark}
\subsection{User Selection Algorithm}
The user selection algorithm is simple:  select the user with the largest $R^i_n$, i.e., select the user with the best rate. The intuition behind this is the following. The rate allocation algorithm of a user $i$ would direct it to transmit at a high rate $R^i_n$ under two circumstances:  either the channel condition for that user is very good, in which case, transmission at high rate saves power, or the delay constraints of that user are not being satisfied. Thus selecting a user with a high rate results in either power savings or the user delay constraint being satisfied. 
\begin{figure}
\centering
\includegraphics[width=3in]{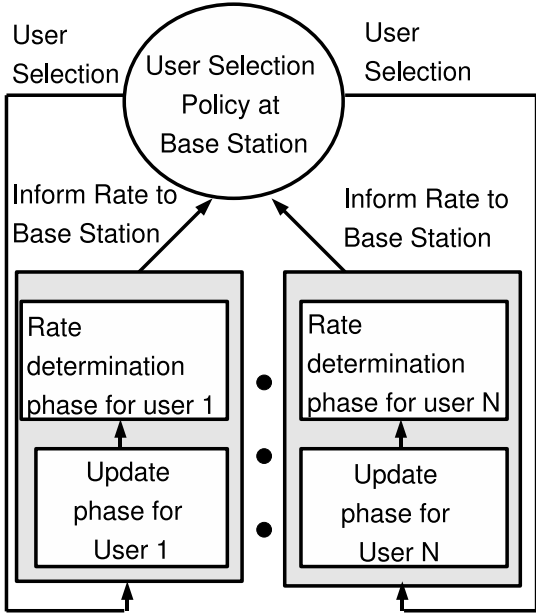}
\caption{Solution schematic}
\label{fig:uplink_schematic}
\end{figure}

% \subsection{Optimality of the Algorithm for $N=1$}
% \begin{theorem}
% The algorithm (\ref{eq:rate_deter}), (\ref{odprvi6}) and (\ref{eq:lm_iter6})  is optimal for $N=1$. 
% \end{theorem}
% \begin{proof}
% For $N=1$, i.e., the single user scenario, the algorithm is same as the algorithm suggested in \cite{jsac_paper}, for which we have proved the optimality in \cite{jsac_paper}.
% \end{proof}

\subsection{Algorithm Details and Implementation}
The rate allocation algorithm is implemented on the user devices while the user selection algorithm is implemented at the base station, as illustrated in Figure \ref{fig:uplink_schematic}. From (\ref{eq:rate_deter}) note that the rate determination phase requires $X^i_{n}$, i.e., the knowledge of the channel state at the base station. The  communication overhead incurred by the base station in informing a user the channel state perceived by it depends on the number of states used to represent the channel. We represent the channel using $8$ states. Hence the base station needs $3$  bits per slot in order to inform a user the channel state perceived by it. The users inform the base station  the rate at which they would transmit if they were to be scheduled. We allocate $3$ bits for conveying this information, i.e., the system can employ $8$ rates. The user selection algorithm then determines the user to be scheduled and all the users are informed about this decision. The rate allocation algorithm at each user then enters the update phase where the value function and the LM for each user are appropriately updated using (\ref{odprvi6}) and (\ref{eq:lm_iter6}). The algorithm thus  continues in each slot $n$. The rate allocation algorithm that is executed at each user device is illustrated in Algorithm \ref{Alg_ud} where, steps 4-8 represent the rate determination phase, while steps 10-14 represent the update phase. The user selection algorithm executed at the base station is detailed in Algorithm \ref{Alg_bs}.

\begin{algorithm}[h]
\caption{The Rate Allocation Algorithm at the User $i$ Device}
\label{Alg_ud}
\begin{algorithmic} [1]
\STATE  Initialize the value function matrix $\tilde{V}^i(q,x) \leftarrow 0 \;\; \forall q \in \mathcal{Q},x \in \mathcal{X}$
\STATE  Initialize LM $\lambda^i \leftarrow 0$
\STATE  Initialize slot counter $n \leftarrow 1$
\STATE  Initialize queue length $q^i \leftarrow 0$
\STATE  Initialize channel states $x^i \leftarrow 0, {x^i}' \leftarrow 0$
\STATE  Reference state $\tilde{s}^{i,0} = (0,x^1)$, where $x^{1} \in \mathcal{X}$
\WHILE  {TRUE}
\WHILE {Base station has not informed the channel state ${x^i}'$}
\STATE wait
\ENDWHILE
\STATE Determine the number of arrivals $A^i_{n+1}=a^i$ in the current slot
\STATE Determine the queue length in the current slot $Q_n^{i}=q^i$
\STATE Use the rate determination phase of the rate allocation algorithm, i.e., (\ref{eq:rate_deter}) to determine the rate $r^i$, for transmission
\STATE Determine the power $P({x^i}',r^i)$ required to transmit $r^i$ bits using (\ref{eq:power1})
\STATE Inform the base station of the rate $r^i$
\WHILE {Base station has not scheduled a user}
\STATE wait
\ENDWHILE
\IF    {user $i$ is scheduled}
\STATE	$u^i \leftarrow r^i$
\ELSE  
\STATE  $u^i \leftarrow 0$
\ENDIF
\STATE  Update the component $(q^i,x^i)$ of the value function matrix $\tilde{V}^i$ using (\ref{odprvi6}). Rest of the components of the matrix remain unchanged
\STATE  Update the LM $\lambda^i$ using (\ref{eq:lm_iter6}) ($Q^i_n=q^i$)
\STATE  $q^i \leftarrow q^i+a^i-u^i$
\STATE  $ x^i \leftarrow {x^i}'$
\label{update_end}
\STATE  $n \leftarrow n+1$
\ENDWHILE
\end{algorithmic}
\end{algorithm}

\begin{algorithm}[h]
\caption{The User Selection Algorithm at the Base Station}
\label{Alg_bs}
\begin{algorithmic} [1]
\WHILE  {TRUE}
\FOR {$i \in 1,\ldots,N$}
\STATE Estimate the channel state $X^i_{n+1}={x^i}'$ in the current slot for user $i$
\STATE Inform ${x^i}'$ to user $i$ 
\ENDFOR
\WHILE {Rate of each user is not known}
\STATE wait
\ENDWHILE
\STATE Determine the user $k$ who has the highest rate
\STATE Schedule user $k$ in the current slot
\ENDWHILE
\end{algorithmic}
\end{algorithm}

\subsection{Discussion}
Here we discuss certain aspects of the online algorithm:
\begin{enumerate}
\item \emph{Computational complexity:} The computational complexity of the rate allocation algorithm executed at a user device is independent of the number of users in the system. This is because the rate allocation algorithm for any user $i$ is directly dependent on the user $i$ state $S^i$ only and is independent of the states of the other users. The user selection algorithm has to determine the maximum of $N$ numbers and hence is linear in $N$. Thus the computational complexity of the user selection algorithm grows only linearly with the number of users.
% \item \emph{Coupling between users:} The heuristic splitting approach does not make the user rate allocation policies completely independent of each other. This is because a user transmitting at a higher rate has a higher chance of being scheduled. If a certain user is scheduled more frequently, the other users would be starved. This would result in their queue lengths growing beyond the queue length constraint, thus forcing them to transmit at higher rates.
\item \emph{An auctioning interpretation:} The solution can be interpreted as an auction, where the user selection algorithm auctions each time slot. The users bid in the form of their transmission rates to the user selection algorithm, which allocates the time slot to the user bidding the highest rate. The rate bid by a user is dependent on its channel state and queue length constraint violation (i.e., the difference between the  current queue length and the queue length constraint). If the channel state is quite good and queue constraint violation is large, the user bids a high rate. This is because transmitting at a high rate when the channel state is good saves power, while doing it when the queue length constraint violation is large aids in satisfying the delays. Note that the users do not bid unnecessarily high rates because that might result in higher power consumption. For a user, not winning an auction in a certain slot, implies that other users either have better channel conditions or higher queue length constraint violation or both. If a user does not win the auction for a certain number of slots successively, its queue length grows thus forcing it to bid a higher rate. Motivated by this interpretation, we refer to the scheduling scheme proposed in this paper as the Auctioning Algorithm (AA).
\end{enumerate}

\section{Simulation Setup and Results}
\label{sec:ssr}

We demonstrate the performance of our algorithm under the IEEE 802.16 \cite{wimax_std} framework through our simulations in a discrete event simulator. Specifically, we intend to demonstrate the following:
\begin{enumerate}
\item The algorithm satisfies the delay constraints of all the users.
\item The algorithm is efficient in terms of the power consumed for each of the users. Moreover, power consumed is commensurate with the delay requirements, average arrival rates and the channel states of the users.  
\item The sum power consumed under our algorithm is marginally more than that consumed under the optimal algorithm in Section \ref{subsec:cmdp}.
\item Average power expended by the users under our algorithm, is much less than that expended under the M-LWDF  scheme \cite{qos_shared}. 
\end{enumerate}
\subsection{M-LWDF Algorithm Details}
The M-LWDF scheduler \cite{qos_shared} attempts to minimize the user delay.  It also considers the probability with which a user's queue length is allowed to exceed a certain target queue length. We assume that this probability is the same for all the users and hence adapt the M-LWDF scheme for our scenario by ignoring it in the present simulations.  Specifically, the adapted M-LWDF scheme schedules a user $i$ in each slot such that,
\begin{equation}
 i = \arg\max_{j} \tau^j_n \times U^j_n,
\end{equation} 
where $\tau^j_n$ is the delay experienced by the head of the line packet for user $j$. 
M-LWDF scheme transmits at a constant maximum power in each time slot. We first determine the average delays experienced by the users under the M-LWDF scheme for various transmission powers. We consider the values of these delays to be the delay constraints for our algorithm.  We determine the average delays experienced by the users under our algorithm and also the power expended by the users under our algorithm for the same maximum transmission power in each slot as in the M-LWDF scheme.  We compare the average power expended by the users under  under our algorithm, with that expended under the M-LWDF  scheme. We perform the simulations within the framework of an IEEE 802.16 system. Next, we provide some details regarding the IEEE 802.16 system.
\subsection{The IEEE 802.16 System}
\label{subsec:wimax}
The IEEE 802.16 standard specifies two modes for sharing the wireless medium: point-to-multipoint (PMP) and mesh. In this paper, we concentrate on the PMP mode where a centralized base station (BS) serves multiple subscriber stations (SSs). We consider the uplink (UL) transmissions. IEEE 802.16 medium access control (MAC) specfies four different scheduling services in order to meet the QoS requirements of various applications. These are: unsolicited grant service (UGS) (for real-time applications with strict delay requirements), real-time polling service (rtPS) (for real-time applications with less stringent delay requirements), non real-time polling service (nrtPS) and best effort (BE) (for applications that do not have any delay requirements). However, unlike BE connection, nrtPS connection is reserved a minimum amount of bandwidth. We consider the \emph{ residential} scenario as in \cite{qos_wimax}. It consists of a BS providing Internet access to the subscribers. Although the standard does not specify any QoS class for providing average delays, we argue that the nrtPS must be extended to cater to the average delay requirements of the users. The unicast polling service of nrtPS can be extended to inform a user the channel state perceived by the base station as well as to determine the rate at which a user would transmit if it were to be scheduled. The scheduling algorithm can thus be implemented as a part of  nrtPS.

The system can be operated in either time division duplex (TDD) or frequency division duplex (FDD) mode. We assume the FDD mode of operation where all SSs have full-duplex capability. We consider a single carrier system (WirelessMAN-SC) with a frame duration of $1$ msec  and bandwidth $W$ of $10$ MHz. We assume that the users transmit at a rate such that data is delivered reliably to the base station. Hence we do not consider retransmissions and Automatic Repeat Request (ARQ).  The SSs employ the following modulations: 64 Quadrature Amplitude Modulation (QAM), 16 QAM, Quadrature Phase Shift Keying (QPSK), QPSK with 1/2 rate convolutional code  which provide us with $4$ rates of transmission. 

\subsection{Simulation Setup and Results}
Internet traffic is modeled as a web traffic source \cite{qos_wimax,m_pareto2}. Packet sizes are drawn from a truncated Pareto distribution (shape factor $1.2$, mode = $2000$ bits, cutoff threshold = $10000$ bits) which provides us with an average packet size of $3860$ bits. In each time frame, we generate the arrivals for all the users using Poisson distribution\footnote{AA does not rely on the Poisson arrival process of the users, we simulate using the Poisson process only for the purpose of illustration.}. Arrivals are generated in an i.i.d. manner across frames. 
We fragment the packets into fragments of size $2000$ bits each. Fragments of size less than $2000$ bits are padded with extra bits to make them of size $2000$ bits. Since all fragments are of equal size, we determine the transmission rate for users in terms of number of fragments per frame instead of bits per frame.
We simulate a Rayleigh fading channel\footnote{AA does not rely on the Rayleigh channel, we simulate using a Rayleigh channel only for the purpose of illustration.} for each user.
For a Rayleigh model, channel state $X^i$ is an
exponentially distributed random variable with probability density
function  given by
$f_{X^i}(x)=\frac{1}{\alpha^i}e^{-\frac{x}{\alpha^i}}$, where $\alpha^i$ is
the mean of $X^i$. 
We know from (\ref{eq:power1})
that the  power required for transmitting $u$ fragments of size $\ell$ bits when the
channel state is $x$ is given by, $P(x,u)=\frac{N_0 W}{x}
\left(2^{u\ell/W}-1 \right)$, where $N_0$ is the power spectral
density of the additive white Gaussian noise and $W$ is the received signal bandwidth. We
assume that the product $WN_{0}$ is normalized to $1$. We measure the sum of queuing and transmission delays of the packets and ignore the propagation delays. In all the scenarios described below, a single simulation run consists of running the algorithm for $100000$ frames and the results are obtained after averaging over $20$ simulation runs.
\begin{scn}
{\em Comparison with the Optimal Algorithm:}
This scenario demonstrates that the sum power expended by AA is very close to that expended by the optimal algorithm (OA) suggested in Section \ref{subsec:cmdp}.  For this scenario, we assume $N=2$, i.e., two users. The channel state can be either bad ($\alpha^{1}$ = $0.1422$ ($-8.47$ dB)) or good ($\alpha^{2}$ = $2.0796$ ($3.18$ dB)). We assume a buffer of  size $10$ packets ($B=10$) at each user. In each frame the arrivals are generated using the Poisson distribution with mean $0.05$ packets/msec ($0.184$ Mbits/sec/user). Packet lengths are Pareto distributed with parameters as discussed previously in this section. In each frame, we generate a Rayleigh random variable with mean $0.9817$ ($-0.08$ dB). If the value taken by the random variable is greater than $2.0796$, the channel state is assumed to be good, else channel state is assumed to be bad. The peak transmission power in any slot is fixed at $3$ Watts. We compare the sum power for the two users for the two schemes in Table \ref{tab:opt_comp}. It can be seen that both the schemes satisfy the delay constraints. The power required by AA  is marginally more than that required for the OA.
\begin{table}
\centering
\caption{Comparison between Optimal Algorithm (OA) and the Auctioning Algorithm (AA)}
\label{tab:opt_comp}
\begin{tabular}{|c|c|c|c|l|} \hline
Delay          & Achieved        & Achieved         & Power for  & Power for \\ 
Constraint   & Delay for OA & Delay for AA   & OA             & AA             \\ \hline
3 msec         & 3.09119         &  3.09700         & 0.26359    & 0.26522\\ \hline
5 msec         & 3.57960         &  3.40470         & 0.24756    & 0.26097 \\ \hline
\hline\end{tabular}
\end{table}
\end{scn}

For the rest of the scenarios, we discretize the channel into eight equal
probability bins, with the boundaries specified by \{ (-$\infty$,
$-8.47$ dB), [$-8.47$ dB, $-5.41$ dB), [$-5.41$ dB, $-3.28$ dB), [$-3.28$ dB,
$-1.59$ dB), [$-1.5$ dB, $-0.08$ dB), [$-0.08$ dB, $1.42$ dB), [$1.42$ dB, $3.18$ dB),
[$3.18$ dB, $\infty$ ) \}. For each bin, we associate a channel state and the state space 
$\mathcal{X}$ = \{ $-13$ dB, $-8.47$ dB, $-5.41$ dB,
$-3.28$ dB, $-1.59$ dB, $-0.08$ dB, $1.42 dB$, $3.18$ dB\}. This discretization of the state space of $X^i$ has
been justified in \cite{wang_mand}. We assume $N=20$, i.e., a system with $20$ users and thereby $20$ UL connections. We assume that the number of users do not change during the course of the simulation. Users are divided into two groups  (Group 1 and Group 2) of $10$ users each. 

\begin{scn}
{\em Comparison with the M-LWDF Algorithm:}
We compare the power consumed by the M-LWDF algorithm with the power consumed by the AA. We first simulate the M-LWDF scheme \cite{qos_shared}. In each frame, arrivals are  generated with a Poisson distribution with mean $0.1$ packets/msec. Packet lengths are Pareto distributed as explained above. This results in an arrival rate of $0.386$ Mbits/sec/user. We choose $\alpha^i = 0.9817 (-0.08\;\mathrm{dB})\;\forall i$. In each slot we generate $X^i$ using exponential distribution with mean $\alpha^i$. We determine the channel state based on the bin that contains $X^i$ as explained above. We perform multiple experiments. In successive experiments, we fix the maximum transmission power at $1.5, 2, 2.5, 3, 4, 4.5$ Watts respectively. For each experiment, we determine the average delays experienced by the users. Next we fix the achieved delays by the M-LWDF algorithm as the delay constraints for our algorithm. The maximum transmission power is the same as that for the M-LWDF algorithm. We determine the average power expended for each user under the AA and compare it to that for the M-LWDF algorithm in Table \ref{tab:mlwdf_comp} (Power constraint and average power expenditure  expressed in Watts, achieved delays expressed in milli-seconds (msec)). It can be seen that for all the experiments, the AA satisfies the delay constraints. Moreover, the power expended by each user under the AA is much less than that expended under the M-LWDF algorithm. This is because the AA is able to adapt the transmission power based on the channel state and delay constraint violation.

\begin{table}
\centering
\caption{Comparison between M-LWDF and the AA}
\label{tab:mlwdf_comp}
\begin{tabular}{|c|c|c|c|l|} \hline
Power          		& Achieved Delay       	& Achieved         & Average Power    & Average   \\ 
Constr.   		&  - M-LWDF 			& Delay - AA      & - M-LWDF       		   & Power - AA \\ \hline
1.5      			&   28.71532  			&  28.12900      &  0.07499   	   & 0.04206  \\ \hline
2         			&   28.18142  			&  28.01677      &  0.09999  	   & 0.04737 \\ \hline
2.5      			&   22.57460       		&  22.03332      &  0.12499   	   & 0.05530   \\ \hline
3         			&   22.12825       		&  18.27730      &  0.14999  	   & 0.07007\\ \hline
3.5      			&   21.99025       		&  16.36487      &  0.17499   	   & 0.07026  \\ \hline
4         			&   20.09445       		&  16.39282      &  0.19999   	   & 0.07073 \\ \hline
4.5      			&   20.09445       		&  14.74980      &  0.22497  	   & 0.07074 \\ \hline
\hline\end{tabular}
\end{table}

\end{scn}

\begin{figure*}
\centerline{\subfigure[Achieved delay of a user]{\includegraphics[width=2.5in,angle=-90]{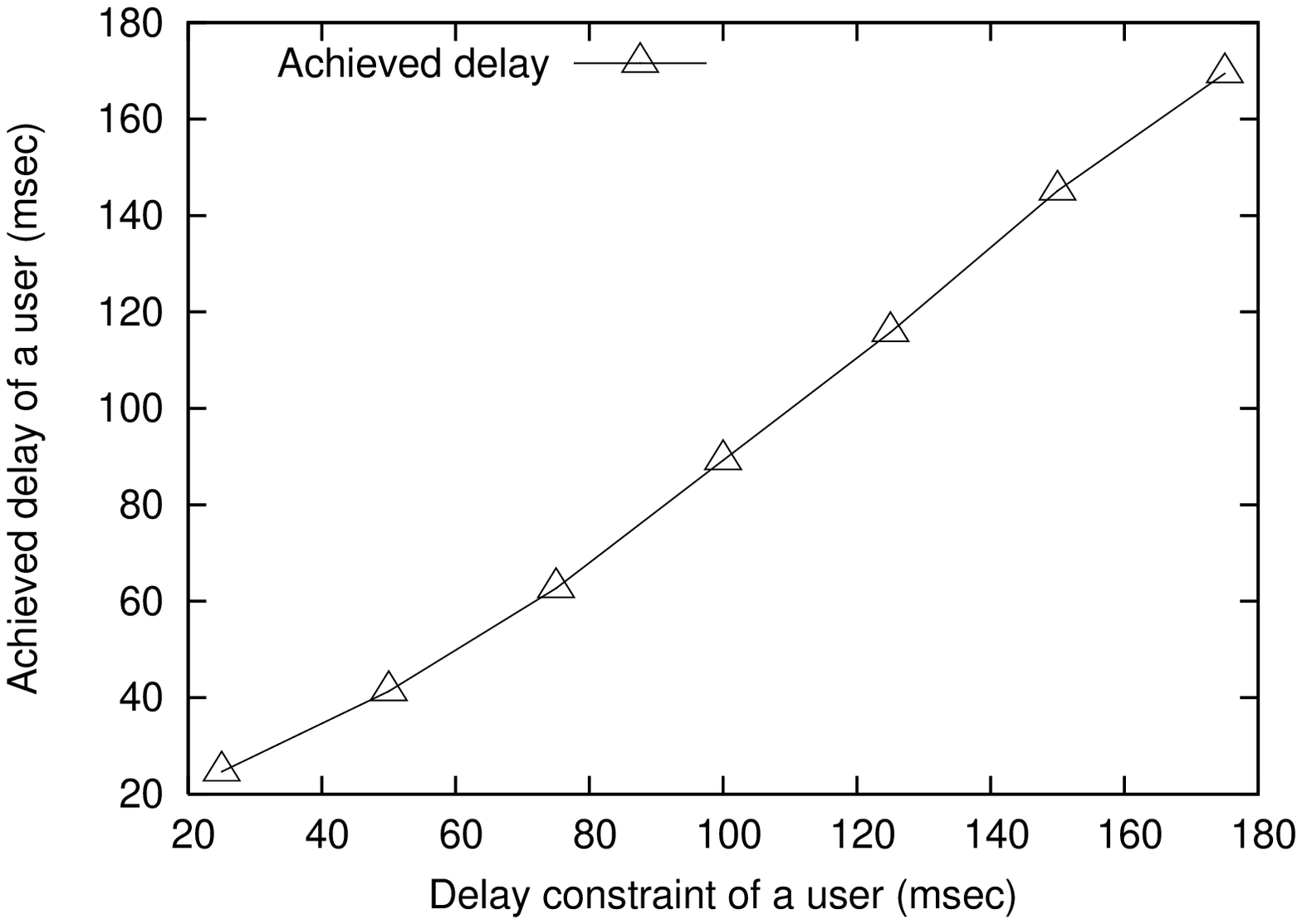}
\label{fig_single_del_delay}}
\hfil
\subfigure[Power consumed by a user]{\includegraphics[width=2.5in,angle=-90]{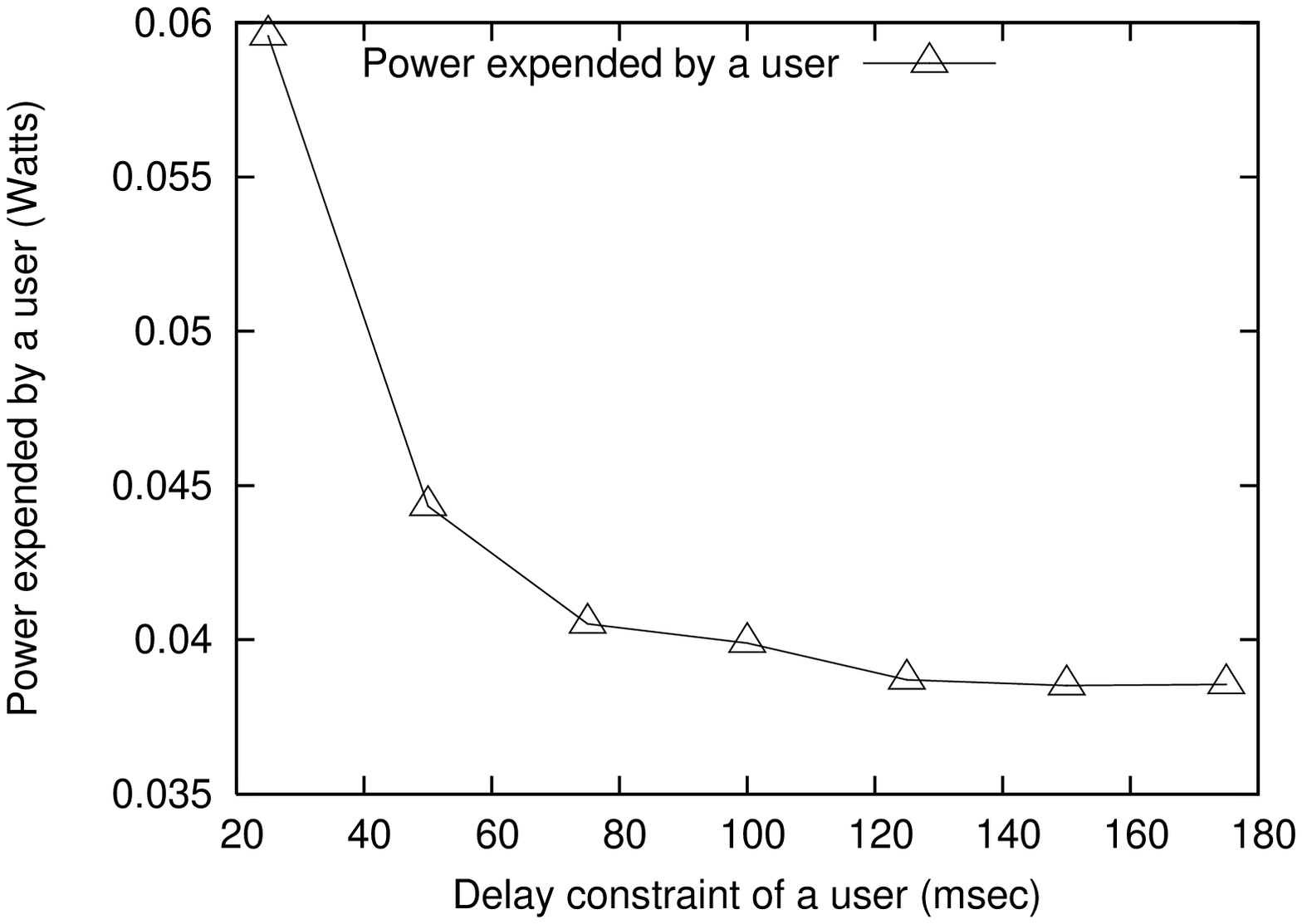}
\label{fig_single_del_power}}}
\caption{Variation of achieved delay and power consumed for various delay constraints - symmetric case}
\label{fig_dv_s}
\end{figure*}

\begin{figure*}
\centerline{\subfigure[Achieved delay of a user]{\includegraphics[width=2.5in,angle=-90]{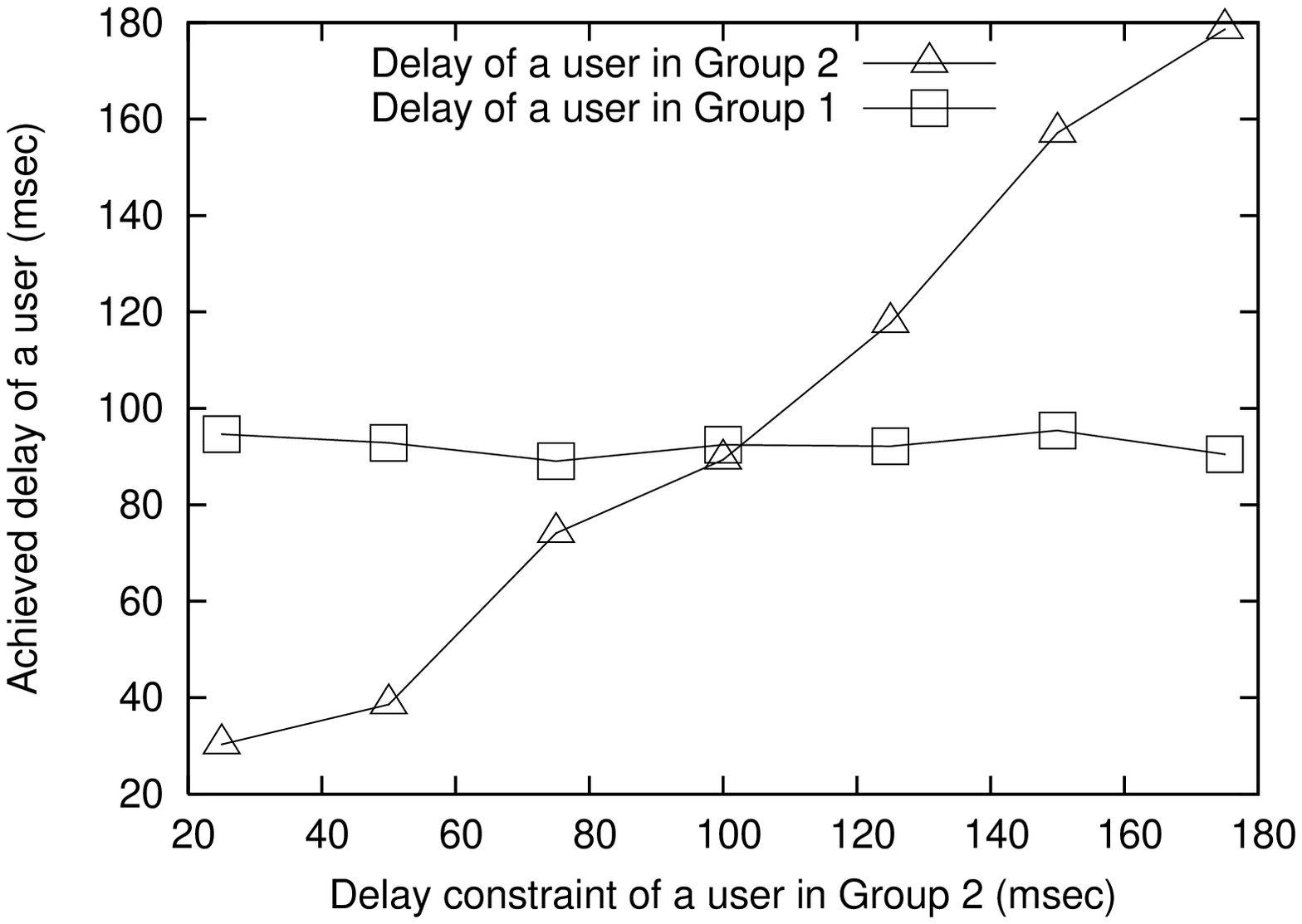}
\label{fig_single_del_delay_asy}}
\hfil
\subfigure[Power consumed by a user]{\includegraphics[width=2.5in,angle=-90]{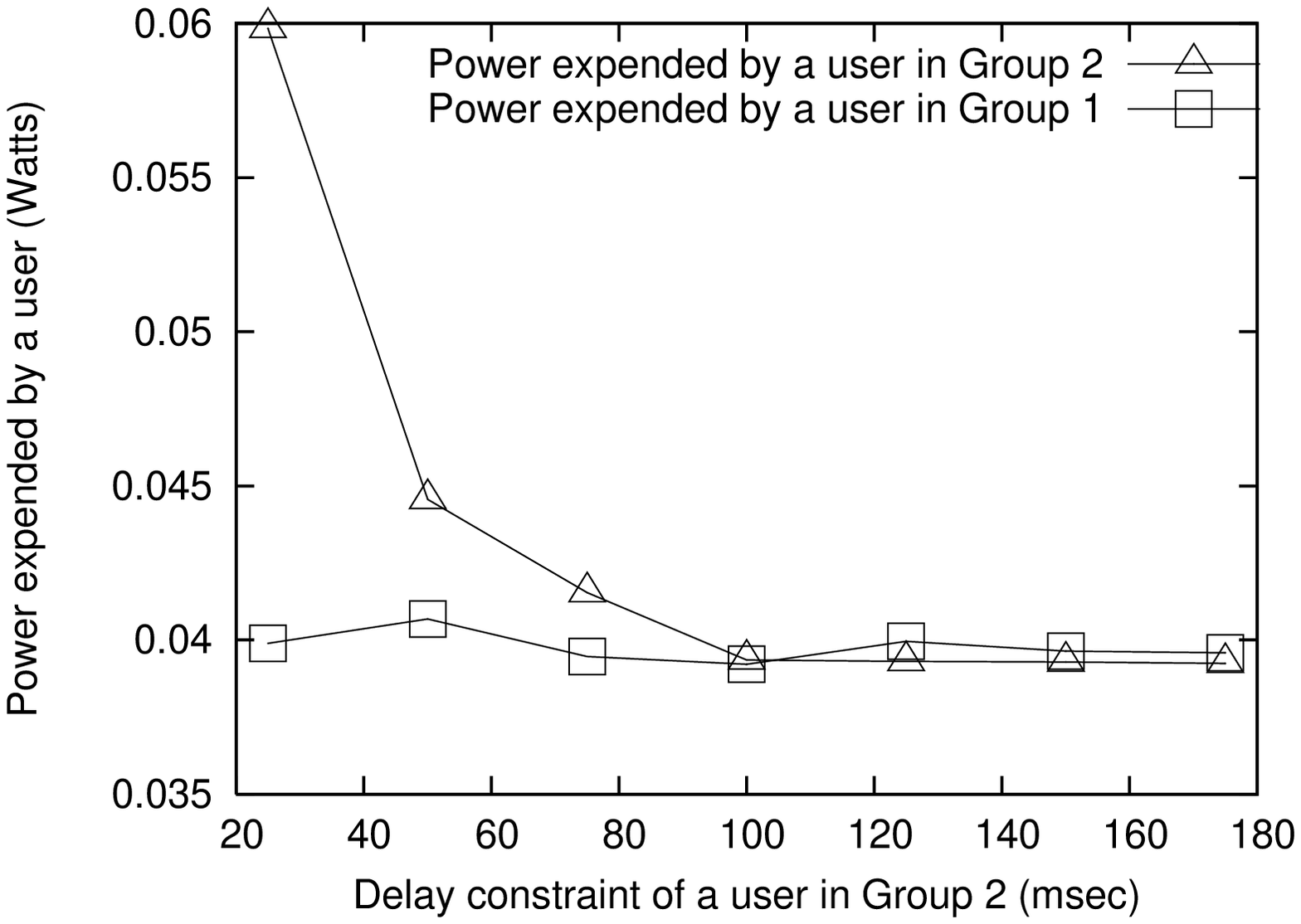}
\label{fig_single_del_power_asy}}}
\caption{Variation of achieved delay and power consumed for various delay constraints - asymmetric case}
\label{fig_dv_as}
\end{figure*}

\begin{scn}
\label{expt:delay}
In this scenario, we demonstrate that the AA satisfies the various user specified delay constraints. We consider two cases: symmetric and asymmetric. In each frame arrivals are generated with a Poisson distribution with mean $0.1$ packets/msec. Packet lengths are Pareto distributed with parameters detailed above. This results in an arrival rate of $0.386$ Mbits/sec/user. We choose $\alpha^i = 0.9817 (-0.08\;\mathrm{dB})\;\forall i$. In each slot we generate $X^i$ using exponential distribution with mean $\alpha^i$. We determine the channel state based on the bin that contains $X^i$ as explained above. 
% We then make use of the rate determination phase of Algorithm \ref{Alg_ud} to determine the rate at which each user transmits if it is scheduled and  the user selection algorithm (\ref{Alg_bs}) to determine the user that is scheduled. The value function matrix and the LMs are then appropriately updated using the update phase of the rate allocation algorithm.
We perform multiple experiments. In the symmetric scenario, in successive experiments, the delay constraints of all the users are fixed at $25, 50, 75, 100, 125, 150, 175$ msec respectively. We measure the average delay experienced and the average power expended by each user in each experiment. These quantities for a user selected at random are plotted in Figure \ref{fig_dv_s}. In the asymmetric case, the delay constraint of the users in Group $1$ are fixed at $100$ msec in each experiment, while the delay constraints of the users in Group $2$ are fixed at $25, 50, 75, 100, 125, 150, 175$ msec in successive experiments. Average delay suffered by a user selected at random from Group 1 and  Group 2 and power consumed by them are plotted in Figure \ref{fig_dv_as}. It can be observed from Figures \ref{fig_single_del_delay} and \ref{fig_single_del_delay_asy} that the delay constraints are satisfied in both the cases. Moreover, from Figures \ref{fig_single_del_power} and \ref{fig_single_del_power_asy} it can be observed that power expended is a convex decreasing function of the delay constraint imposed by the user. Larger delay constraints imply that much lesser power is required to satisfy the constraint.
\end{scn}

\begin{figure*}
\centerline{\subfigure[Achieved delay of a user]{\includegraphics[width=2.5in,angle=-90]{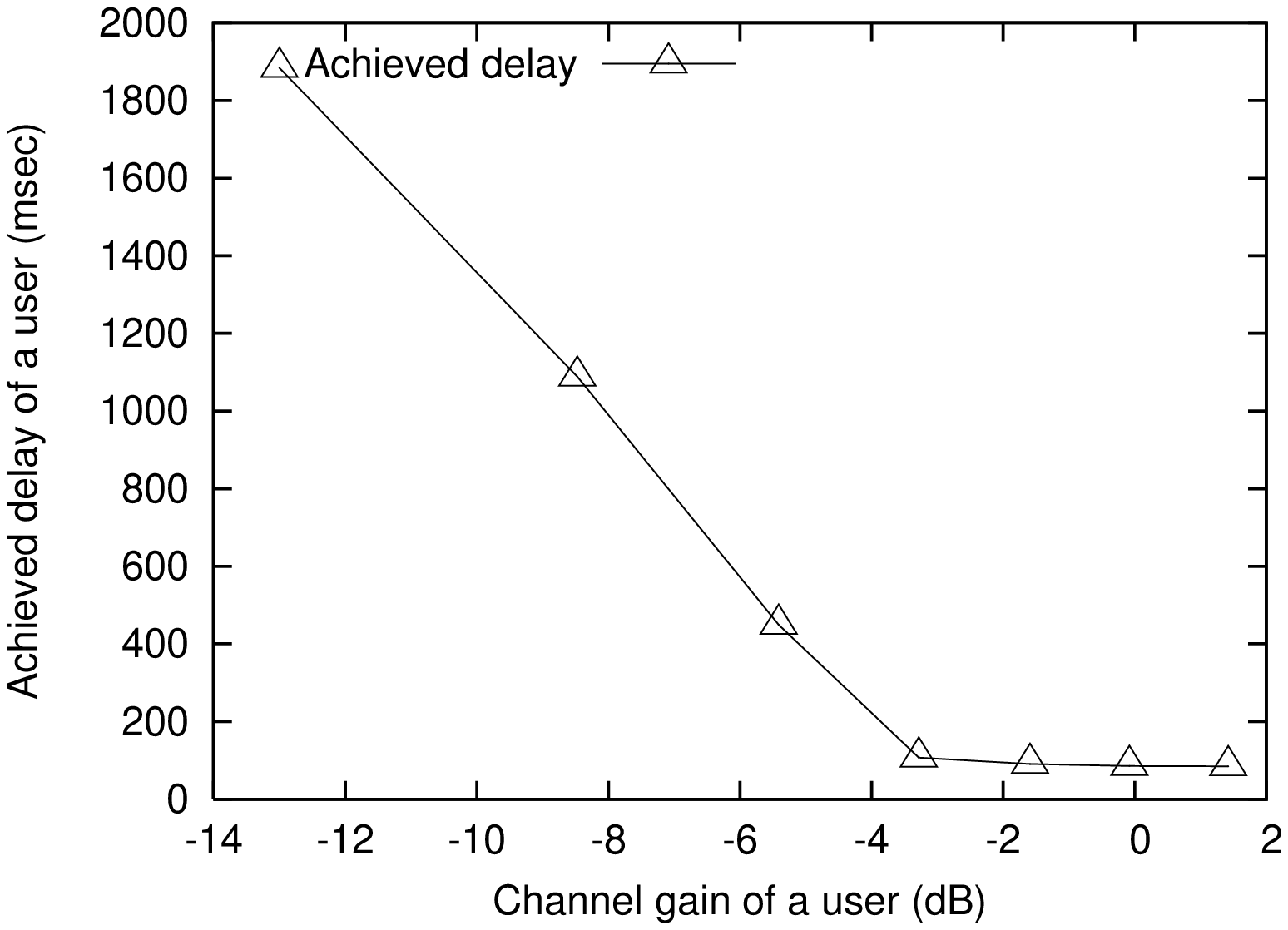}
\label{fig_single_ch_delay}}
\hfil
\subfigure[Power consumed by a user]{\includegraphics[width=2.5in,angle=-90]{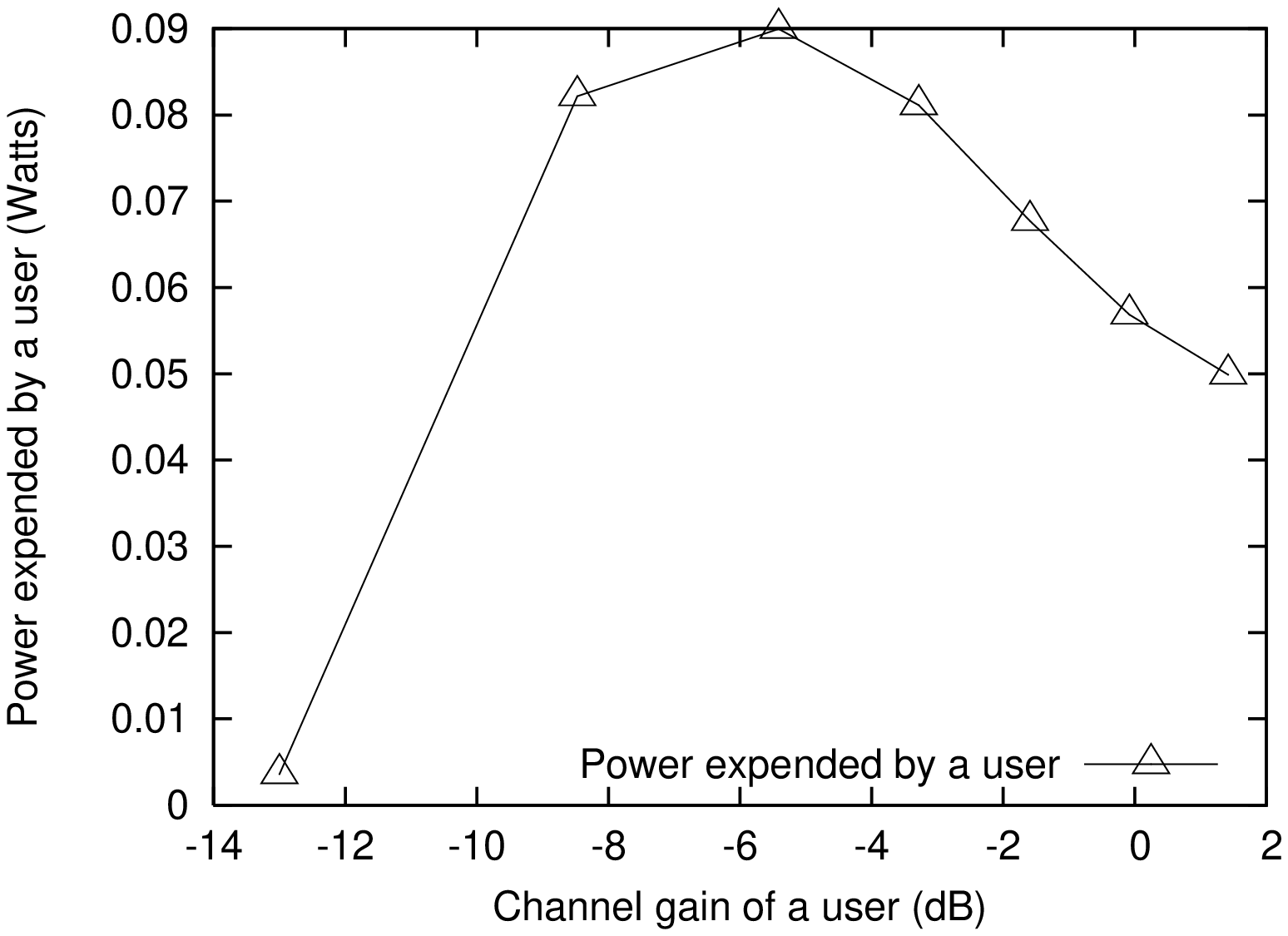}
\label{fig_single_ch_power}}}
\caption{Variation of achieved delay and power consumed for various channel conditions - symmetric case}
\label{fig_cv_s}
\end{figure*}

\begin{figure*}
\centerline{\subfigure[Achieved delay of a user]{\includegraphics[width=2.5in,angle=-90]{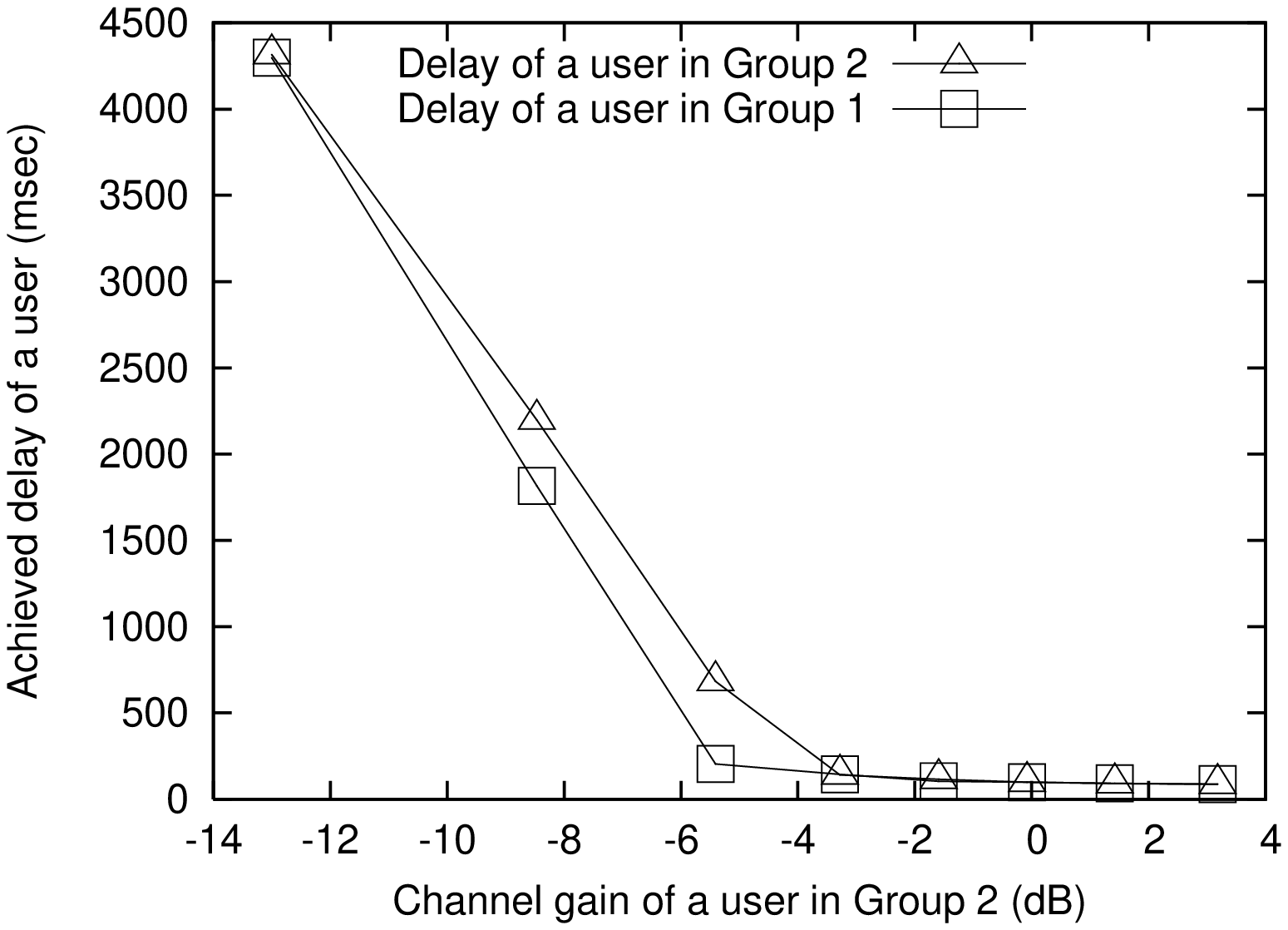}
\label{fig_single_ch_delay_asy}}
\hfil
\subfigure[Power consumed by a user]{\includegraphics[width=2.5in,angle=-90]{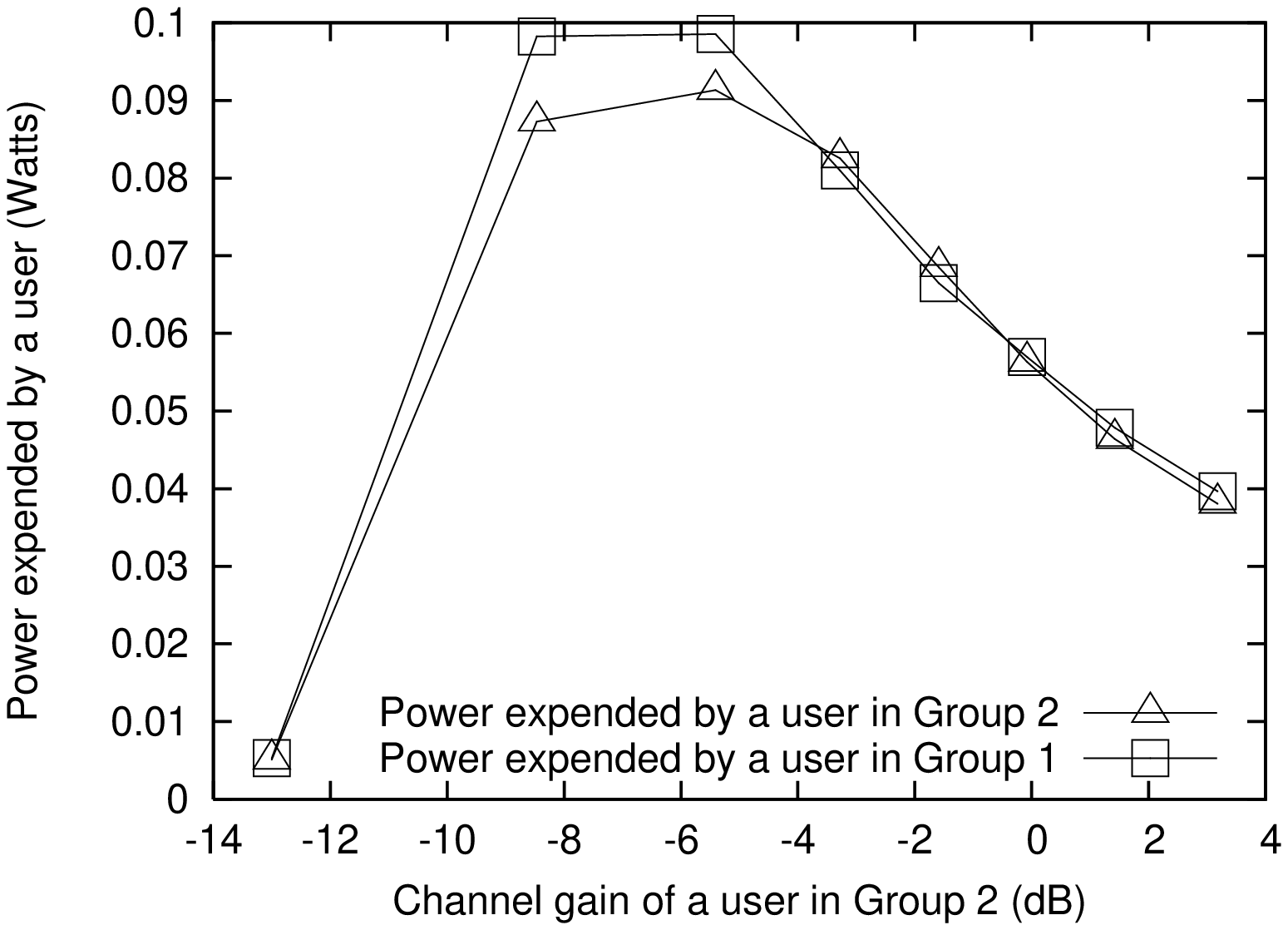}
\label{fig_single_ch_power_asy}}}
\caption{Variation of achieved delay and power consumed for various channel conditions - asymmetric case}
\label{fig_cv_as}
\end{figure*}

\begin{scn}
In this scenario, we demonstrate that the AA satisfies the user specified delay constraints for various channel conditions. We consider two cases: symmetric and asymmetric. The delay constraints of all the users are kept constant at $100$ msec. For the symmetric case, we fix  $\alpha^i$ as $-13\;\mathrm{dB}$, $-8.47\;\mathrm{dB}$, $-5.41\;\mathrm{dB}$, $-3.28\;\mathrm{dB}$, $-1.59\;\mathrm{dB}$, $-0.08\;\mathrm{dB}$, $1.42\;\mathrm{dB}$, $\forall i$ in successive experiments.  Rest of the parameters are the same as in Scenario \ref{expt:delay}. We measure the average delay suffered by each of the users and the average power consumed by each of them. These quantities are plotted in Figure \ref{fig_cv_s}. In the asymmetric case, we maintain the average channel state for users in Group 1 constant for all the experiments, i.e., $\alpha^i = -0.08\;\mathrm{dB}, \;i \in 1,\ldots,10$. For the users in Group 2,  i.e., $\alpha^i$ for $i \in 11,\ldots,20$,   the average channel state is fixed at $\alpha^i$ $=$ $-13\;\mathrm{dB}$, $-8.47\;\mathrm{dB}$, $-5.41\;\mathrm{dB}$, $-3.28\;\mathrm{dB}$, $-1.59\;\mathrm{dB}$, $-0.08\;\mathrm{dB}$, $1.42\;\mathrm{dB}$,  in successive experiments. Average delay suffered by a user in Group 1 and in Group 2  and power consumed by them are plotted in Figure \ref{fig_cv_as}. It can be observed from Figures \ref{fig_single_ch_delay} and \ref{fig_single_ch_delay_asy} that the delay constraints are satisfied even for extremely poor channel conditions. Moreover, from Figures \ref{fig_single_ch_power} and \ref{fig_single_ch_power_asy} it can be observed that the scheme is able to satisfy the delay constraints above a certain average channel state\footnote{This average channel state is dependent on the peak transmission power.}. Better channel conditions imply that much lesser power is required to satisfy the delay constraints.
\end{scn}

\begin{figure*}
\centerline{\subfigure[Achieved delay of a user]{\includegraphics[width=2.5in,angle=-90]{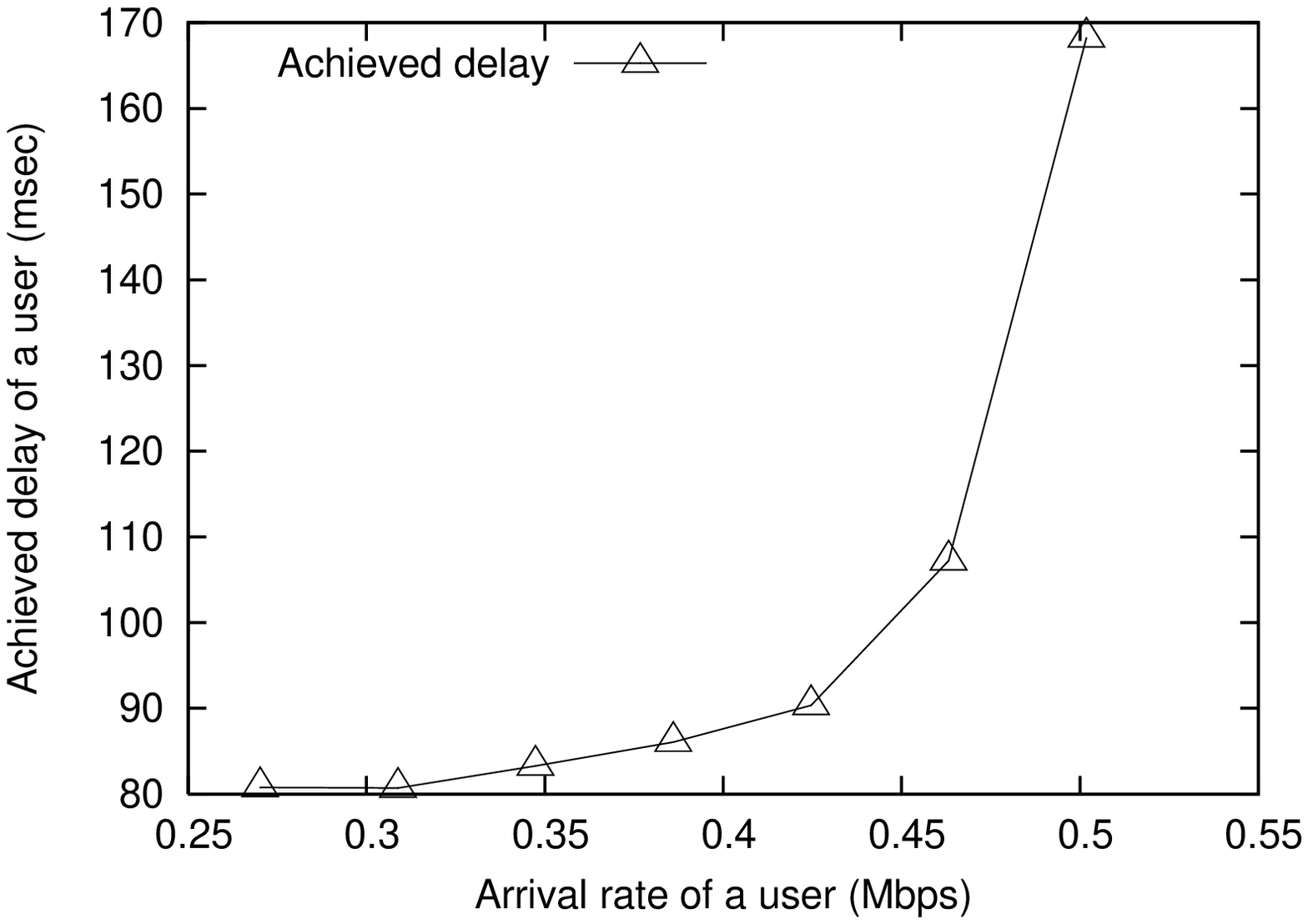}
\label{fig_single_ar_delay}}
\hfil
\subfigure[Power consumed by a user]{\includegraphics[width=2.5in,angle=-90]{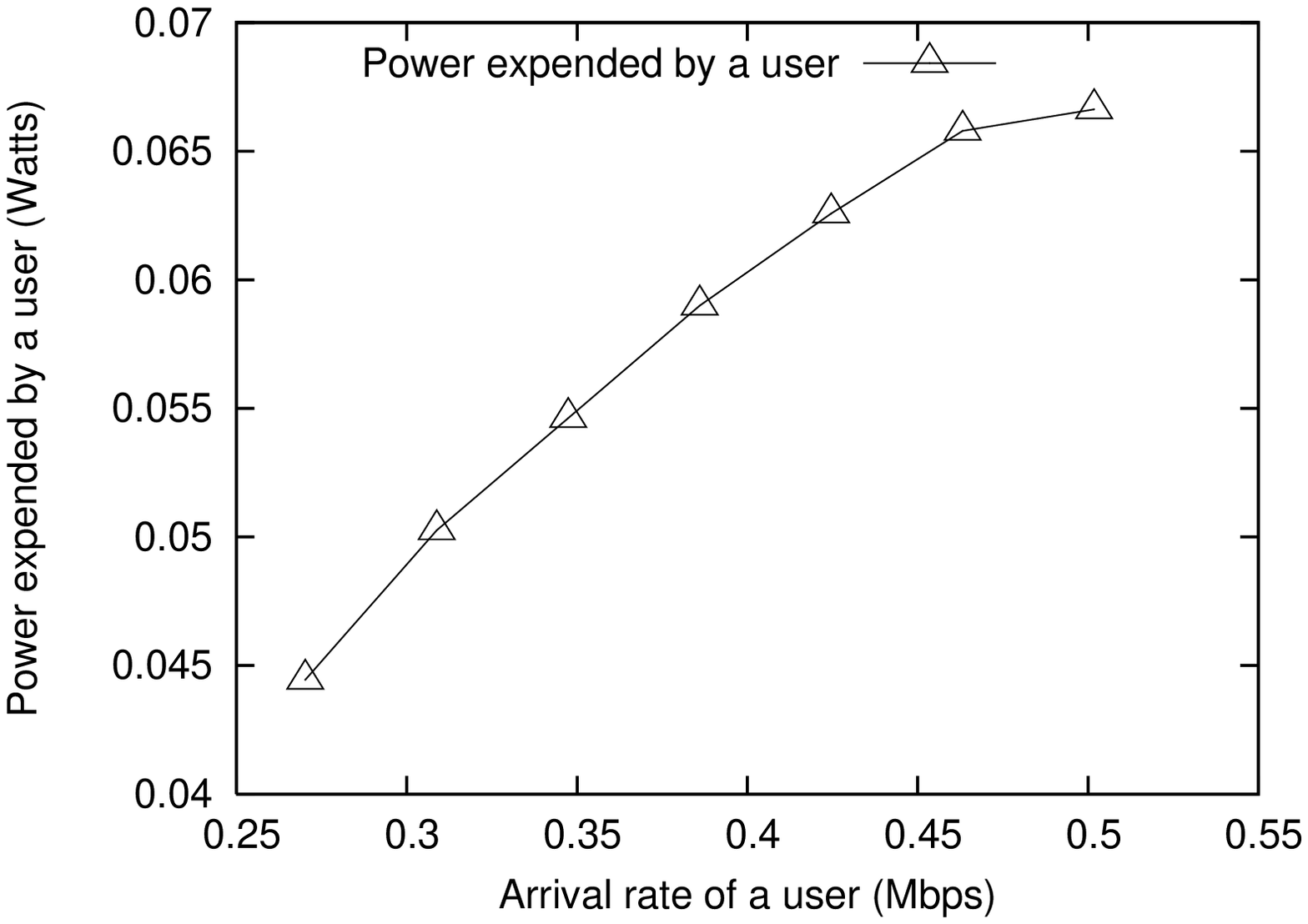}
\label{fig_single_ar_power}}}
\caption{Variation of achieved delay and power consumed for various arrival rates - symmetric case}
\label{fig_av_s}
\end{figure*}

\begin{figure*}
\centerline{\subfigure[Achieved delay of a user]{\includegraphics[width=2.5in,angle=-90]{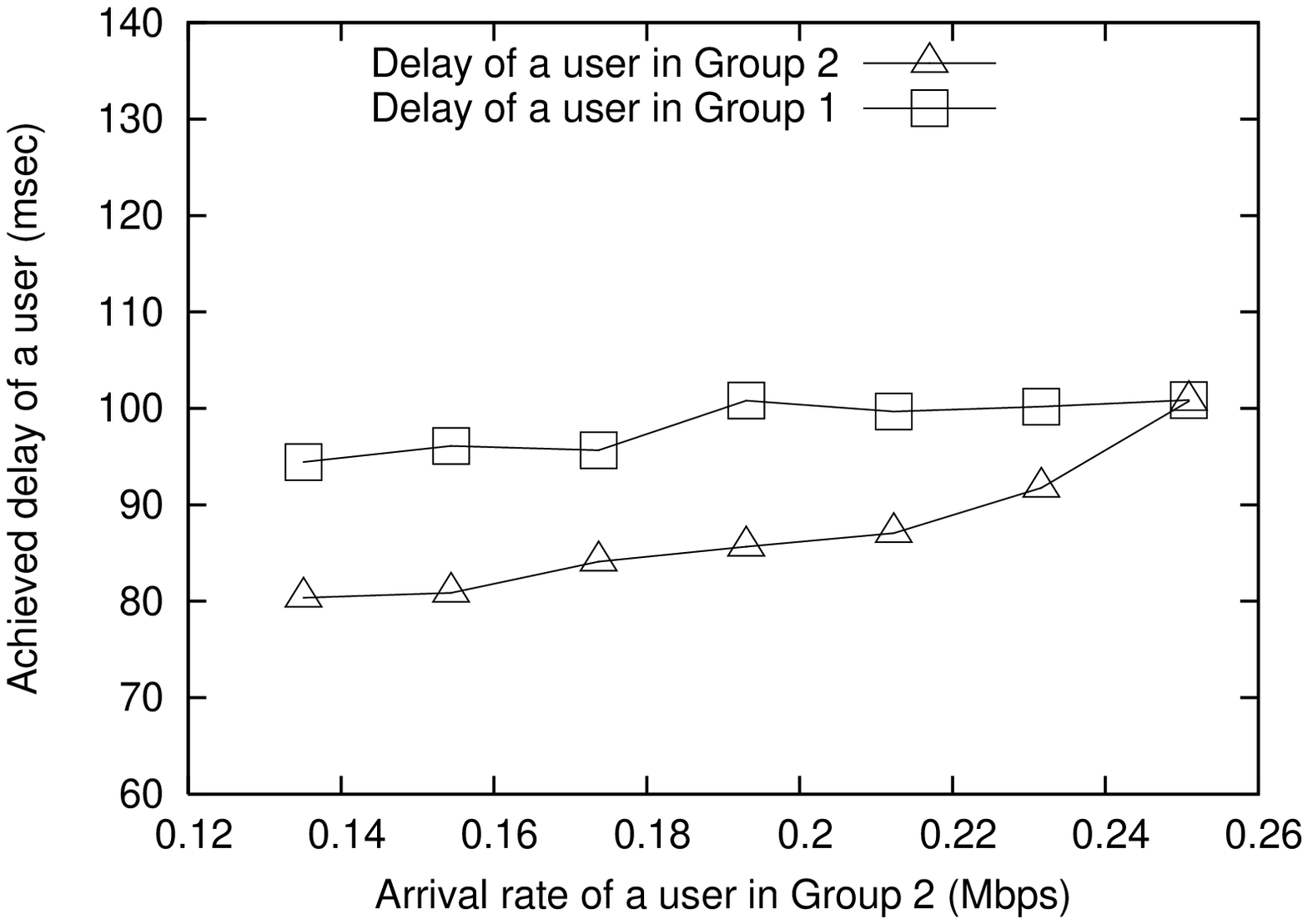}
\label{fig_single_ar_delay_asy}}
\hfil
\subfigure[Power consumed by a user]{\includegraphics[width=2.5in,angle=-90]{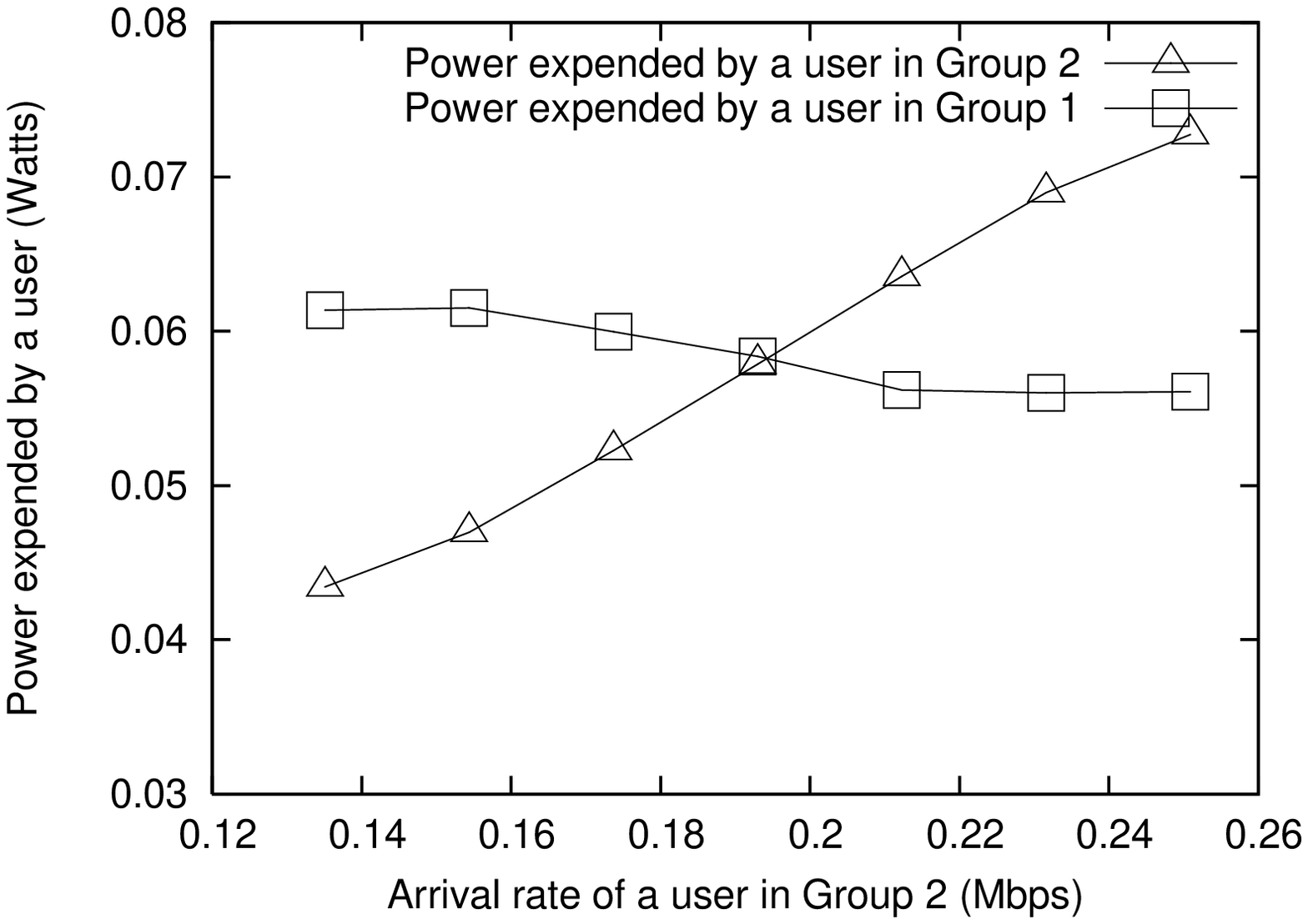}
\label{fig_single_ar_power_asy}}}
\caption{Variation of achieved delay and power consumed for various arrival rates - asymmetric case}
\label{fig_av_as}
\end{figure*}

\begin{scn}
In this scenario we demonstrate the range of arrival rates for which the AA satisfies the user specified delay constraint of $100$ msec. We consider two cases - symmetric and asymmetric. In the symmetric case, the arrival rates of all the users are fixed at $0.2702 \mathrm{to} 0.5018$ Mbits/sec ($0.05 \mathrm{to}0.12$ packets/msec) in successive experiments. Rest of the parameters are same as in Scenario \ref{expt:delay}. We measure the average delay suffered and average power expended by each user. These quantities for a user chosen at random are plotted in Figure \ref{fig_av_s}. In the asymmetric case, the arrival rate of the users in Group 1 is fixed at $0.386$ Mbits/sec ($0.15$ packets/msec) for all the experiments, while the arrival rates of the users in Group 2 are increased from $0.1351-0.2509$ Mbits/sec ($0.07-0.13$ packets/msec) in $8$ steps in successive experiments. Rest of the parameters are same as in Scenario \ref{expt:delay}. Average delay suffered by a user from Group1 and Group 2 (each selected at random) and power consumed by them are plotted in Figure \ref{fig_av_as}. It can be observed from Figures \ref{fig_single_ar_delay} and \ref{fig_single_ar_delay_asy} that the delay constraints are satisfied in both the cases. From Figures \ref{fig_single_ar_power} and \ref{fig_single_ar_power_asy} it can be seen that power expended is an  increasing function of the average arrival rates for the same delay constraint. Higher the arrival rate, higher is the power expended.
\end{scn}

% \vspace{-1mm}
\section{Conclusion}
\label{sec:concl}
In this paper, we have considered the problem of power efficient uplink scheduling in a TDMA system over a fading wireless  channel with the objective of minimizing user power expenditure  under individual delay constraints. We have assumed that the user statistics are unknown, i.e., the probability distributions of the user arrivals and channel states are unknown. We have formulated the problem under the CMDP framework. Determining the optimal policy under the CMDP framework faces two problems: state space explosion and unknown user statistics. To tackle state space explosion, we have suggested performing  the rate allocation  for a particular user  based on its buffer occupancy and channel state only. The rate allocation algorithm is a learning algorithm that learns about the channel state and buffer occupancy of a user during system execution and determines its rate of transmission and hence takes care of the unknown user statistics. Once the rate allocation for all the users is is done, the algorithm schedules a user with the highest rate in a slot. We have performed simulations in  the IEEE 802.16 system. Our simulation results have demonstrated that the system is indeed able to satisfy the user specified delay constraints and comparison with the M-LWDF scheme indicates that power expended by the users  is  low.  Moreover, the power expended is commensurate with the QoS requirement, lower average arrival rates, better average channel conditions and higher average delay requirements lead to lower power expenditure.

\bibliographystyle{IEEEtran.bst}

\bibliography{tmc}

\begin{thebibliography}{10}
\providecommand{\url}[1]{#1}
\csname url@rmstyle\endcsname
\providecommand{\newblock}{\relax}
\providecommand{\bibinfo}[2]{#2}
\providecommand\BIBentrySTDinterwordspacing{\spaceskip=0pt\relax}
\providecommand\BIBentryALTinterwordstretchfactor{4}
\providecommand\BIBentryALTinterwordspacing{\spaceskip=\fontdimen2\font plus
\BIBentryALTinterwordstretchfactor\fontdimen3\font minus
  \fontdimen4\font\relax}
\providecommand\BIBforeignlanguage[2]{{%
\expandafter\ifx\csname l@#1\endcsname\relax
\typeout{** WARNING: IEEEtran.bst: No hyphenation pattern has been}%
\typeout{** loaded for the language `#1'. Using the pattern for}%
\typeout{** the default language instead.}%
\else
\language=\csname l@#1\endcsname
\fi
#2}}

\bibitem{wimax_std}
{LAN/MAN Standards Committee}, \emph{{IEEE standard for Local and Metropolitan
  Area Networks - Part 16: Air Interface for Fixed Broadband Wireless Access
  Systems (IEEE P802.16-REVd/D5)}}.\hskip 1em plus 0.5em minus 0.4em\relax IEEE
  Computer Society, May 2004.

\bibitem{3gpp_site}
\BIBentryALTinterwordspacing
``{Third Generation Partnership Project}.'' [Online]. Available:
  \url{http://www.3gpp.org/}
\BIBentrySTDinterwordspacing

\bibitem{tse_book}
D.~Tse and P.~Viswanath, \emph{Fundamentals of Wireless Communication}.\hskip
  1em plus 0.5em minus 0.4em\relax Cambridge University Press, 2005.

\bibitem{tse_hanly1}
D.~Tse and S.~Hanly, ``{Multi-access Fadaing Channels - Part I: Polymatroid
  Structure, Optimal Resource Allocation and Throughput Capacities},''
  \emph{{IEEE} Transactions on Information Theory}, vol.~44, no.~7, pp.
  2796--2815, 1998.

\bibitem{knopp}
R.~Knopp and P.~A. Humblet, ``{Information Capacity and Power Control in
  Single-Cell Multiuser Communications},'' in \emph{Proceedings of {IEEE}
  {ICC}}, vol.~1, Seattle, USA, June 1995, pp. 331--335.

\bibitem{cross_layer_review}
S.~Shakkottai, T.~S. Rappaport, and P.~C. Karlsson, ``{Cross-Layer Design for
  Wireless Networks},'' \emph{{IEEE} Communications Magazine}, vol.~41, no.~10,
  pp. 74--80, 2003.

\bibitem{berryreview}
R.~Berry and E.~M. Yeh, ``{Cross-Layer Wireless Resource Allocation},''
  \emph{{IEEE} Signal Processing Magazine}, vol.~21, no.~5, pp. 59--68, 2004.

\bibitem{yeh8}
E.~Yeh and A.~Cohen, ``{Throughput and Delay Optimal Resource Allocation in
  Multiaccess Fading Channels},'' in \emph{Proceedings of International
  Symposium on Information Theory}, 2003, p. 245.

\bibitem{neely1}
M.~J. Neely, E.~Modiano, and C.~Rohrs, ``{Power and Server Allocation in
  Multi-beam Satellite with Time Varying Channels},'' in \emph{Proceedings of
  {IEEE} {INFOCOM}}, San Francisco, USA, 2002, pp. 138--152.

\bibitem{tass_lcq}
L.~Tassiulas and A.~Ephremides, ``{Dynamic Server Allocation to Parallel Queues
  with Randomly Varying Connectivity},'' \emph{{IEEE} Transactions on
  Information Theory}, no.~2, pp. 466--478, 1993.

\bibitem{stolyar}
S.~Shakkottai and A.~Stolyar, ``{Scheduling for Multiple Fows Sharing a
  Time-varying Channel: the Exponential Rule},'' \emph{{AMS Translations Series
  2}}, vol. 207, 2002.

\bibitem{yeh1}
E.~Yeh and A.~Cohen, ``{Delay Optimal Rate Allocation in Multiaccess Fading
  Communications},'' in \emph{Proceedings of {IEEE} Workshop on Multimedia
  Signal Processing}, Oct. 2002, pp. 404--407.

\bibitem{qos_shared}
M.~Andrews, K.~Kumaran, K.~Ramanan, A.~Stolyar, P.~Whiting, and R.~Vijayakumar,
  ``{Providing Quality of Service over a Shared Wireless Link},'' \emph{{IEEE
  Communications Magazine}}, vol.~39, Feb. 1996.

\bibitem{tse_hanly2}
D.~Tse and S.~Hanly, ``{Multi-access Fadaing Channels - Part II: Delay-Limited
  Capacities},'' \emph{{IEEE} Transactions on Information Theory}, vol.~44,
  no.~7, pp. 2816--2831, 1998.

\bibitem{yeh_thesis}
E.~Yeh, ``{Multiaccess and Fading in Communication Networks},'' {PhD} Thesis,
  Massachusetts Institute of Technology, September 2001.

\bibitem{tse_channel}
E.~F. Chaponniere, P.~Black, J.~M. Holtzman, and D.~Tse, ``{Transmitter
  directed Multiple Receiver System using Path Diversity to Equitably Maximize
  Throughput },'' Sept. 2002, {U.S. Patent No. 6449490}.

\bibitem{liu_shroff2}
X.~Liu, E.~Chong, and N.~Shroff, ``{Opportunistic Transmission Scheduling with
  Resource-Sharing Constraints in Wireless Networks},'' \emph{{IEEE Journal on
  Selected Areas in Communication}}, vol.~19, no.~10, pp. 2053--2064, 2001.

\bibitem{berry}
R.~A. Berry and R.~G. Gallager, ``{Communication over Fading Channels with
  Delay Constraints},'' \emph{{IEEE} Transactions on Information Theory},
  vol.~48, no.~5, pp. 1135--1149, 2002.

\bibitem{wang_mand}
H.~Wang and N.~Mandayam, ``{A Simple Packet Transmission Scheme for Wireless
  Data over Fading Channels},'' \emph{{IEEE Transcations on Communications}},
  vol.~52, no.~7, pp. 1055--1059, 2004.

\bibitem{goyal}
M.~Goyal, A.~Kumar, and V.~Sharma, ``{Power Constrained and Delay Optimal
  Policies for Scheduling Transmissions over a Fading Channel},'' in
  \emph{Proceedings of {IEEE} {INFOCOM}}, vol.~1, Mar. 2003, pp. 311--320.

\bibitem{mukul}
\BIBentryALTinterwordspacing
M.~Agarwal, A.~Karandikar, and V.~S. Borkar, ``{Structural Properties of
  Optimal Transmission Policies over a Randomly Varying channel},'' \emph{Under
  review in {IEEE} Transactions on Automatic Control}, 2006. [Online].
  Available: \url{http://www.ee.iitb.ac.in/~karandi/pubs_dir/mukul.ps.gz}
\BIBentrySTDinterwordspacing

\bibitem{krish}
D.~V. Djonin and V.~Krishnamurthy, ``{Structural Results on the Optimal
  Transmission Scheduling Policies and Costs for Correlated Sources and
  Channels},'' in \emph{{IEEE CDC}}, Dec. 2005, pp. 3231--3236.

\bibitem{balaji}
B.~Prabhakar, E.~U. Biyikoglu, and A.~E. Gamal, ``{Energy-Efficient
  Transmission over a Wireless Link via Lazy Packet Scheduling},'' in
  \emph{Proceedings of {IEEE} {INFOCOM}}, vol.~1, Apr. 2001, pp. 386--394.

\bibitem{cruz_dp}
B.~E. Collins and R.~L. Cruz, ``{Transmission Policies for Time Varying
  Channels with Average Delay Constraints},'' in \emph{Proceedings of 1999
  Allerton Conference on Communication, Control, and Computation}, 1999.

\bibitem{gauresh}
G.~Rajadhyaksha and V.~S. Borkar, ``{Transmission Rate Control over Randomly
  Varying Channels},'' \emph{{Probability in the Engineering and Informational
  Sciences}}, vol.~19, no.~1, pp. 73--82, Jan. 2005.

\bibitem{neely2}
M.~J. Neely, ``{Optimal Energy and Delay Tradeoffs for Multi-user Wireless
  Downlinks},'' in \emph{Proceedings of {IEEE} {INFOCOM}}, 2006, pp. 1--13.

\bibitem{berry_thesis}
R.~Berry, ``{Power and Delay Trade-offs in Fading Channels},'' June 2000, {PhD}
  Thesis, Massachusetts Institute of Technology.

\bibitem{jsac_paper}
\BIBentryALTinterwordspacing
N.~Salodkar, A.~Bhorkar, A.~Karandikar, and V.~S. Borkar, ``{On-Line Learning
  Algorithm for Energy Efficient Delay Constrained Scheduling over Fading
  Channel},'' May 2007, {Submitted}. [Online]. Available:
  \url{http://www.it.iitb.ac.in/~nitins/post_decision-1.pdf}
\BIBentrySTDinterwordspacing

\bibitem{chang1}
H.~S. Wang and N.~Moayeri, ``{Finite-State Markov Channel -- A Useful Model for
  Radio Communication Channels},'' \emph{{IEEE Transactions on Vehicular
  Technology}}, vol.~44, no.~1, pp. 163--171, Feb. 1995.

\bibitem{gallager}
D.~P. Bertsekas and R.~Gallager, \emph{Data Networks}.\hskip 1em plus 0.5em
  minus 0.4em\relax Prentice Hall, 1987.

\bibitem{multi_objective}
C.~A.~C. Coello, ``{A Comprehensive Survey of Evolutionary-Based Multiobjective
  Optimization Techniques},'' \emph{{Knowledge and Information Systems}},
  vol.~3, no.~1, pp. 269--308, 1999.

\bibitem{altman}
E.~Altman, \emph{Constrained Markov Decision Processes}.\hskip 1em plus 0.5em
  minus 0.4em\relax Boca Raton, FL: Chapman and Hall/{CRC} Press, 1999.

\bibitem{ber}
D.~P. Bertsekas, \emph{Dynamic Programming and Optimal Control}.\hskip 1em plus
  0.5em minus 0.4em\relax Belmont, MA: Athena Scientific, 1995, vol.~1.

\bibitem{bor_two_time}
V.~S. Borkar, ``{Stochastic Approximation with Two Time Scales},''
  \emph{Systems and Control Letters}, vol.~29, pp. 291--294, 1996.

\bibitem{puterman}
M.~Puterman, \emph{Markov Decision Processes}.\hskip 1em plus 0.5em minus
  0.4em\relax New York: Wiley, 1994.

\bibitem{young_book}
H.~P. Young, \emph{Strategic Learning and its Limits}.\hskip 1em plus 0.5em
  minus 0.4em\relax New York: Oxford University Press, 2004.

\bibitem{young_poss}
------, ``{The Possible and the Impossible in Multi-Agent Learning},''
  \emph{Artificial Intelligence}, vol. 171, no.~7, pp. 429--433, 2007.

\bibitem{qos_wimax}
C.~Cicconeti, L.~Lenzini, E.~Mingozzi, and C.~Eklund, ``{Quality of Service
  Support in IEEE 802.16 Networks},'' \emph{IEEE Network}, pp. 50--55,
  March/April 2006.

\bibitem{m_pareto2}
T.~G. Neame, M.~Zukerman, and R.~G. Addie, ``{Modeling Broadband Traffic
  Streams},'' in \emph{Proceedings of {IEEE} {GLOBECOM}}, 1999, pp. 1048--1052.

\end{thebibliography}

% \newpage

\end{document}